\documentclass[aps,prb,twocolumn,amsmath,amssymb,superscriptaddress,scrartcl,eqsecnum,longbibliography,nofootinbib]{revtex4-2}

\usepackage{xcolor}
\usepackage{graphicx}
\usepackage{subfigure}
\usepackage{amsthm}
\usepackage{mathtools}
\usepackage{braket}

\usepackage{tikz}
\usetikzlibrary{decorations.pathmorphing}

\makeatletter
\PassOptionsToPackage{caption=false}{subfig} 
\usepackage{hyperref}
\hypersetup{
breaklinks=true,
colorlinks=true,
citecolor=blue,
linkcolor=black,
filecolor=black,
urlcolor=blue
}
\IfFileExists{lmodern.sty}{\usepackage{lmodern}}{}

\def\be{\begin{equation}}
\def\ee{\end{equation}}

\def\Gr{\mathbf{Gr}}

\def\vac{|{\rm vac} \rangle}

\renewcommand{\bar}{\overline}

\usepackage{nicematrix}

\DeclareMathOperator{\Tr}{Tr}
\DeclareMathOperator{\sgn}{sgn}

\newcommand{\norm}[1]{\left\lVert#1\right\rVert}

\newtheorem{theorem}{Theorem}

\newtheorem{proposition}{Proposition}

\newtheorem{lemma}{Lemma}[theorem]

\newtheorem{applemma}{Lemma}[section]
\newtheorem{appprop}{Proposition}[section]

\begin{document}


\title{Universal and Maximal Entanglement Swapping in\\
General Fermionic Gaussian States}

\author{Jiyuan Fang}
\thanks{These two authors contributed equally.}
\affiliation{School of Physics, Georgia Institute of Technology, Atlanta, GA 30332, USA}

\author{Qicheng Tang}
\thanks{These two authors contributed equally.}
\affiliation{School of Physics, Georgia Institute of Technology, Atlanta, GA 30332, USA}

\author{Xueda Wen}
\affiliation{School of Physics, Georgia Institute of Technology, Atlanta, GA 30332, USA}

\begin{abstract}
Exploring universal entanglement structure in many-body systems is both fundamental and challenging, particularly when the system undergoes non-unitary operations.
In this work, we uncover a universal mechanism for realizing maximal entanglement swapping in fermionic Gaussian states subjected to projective Bell measurements.
 We consider two initially decoupled, half-filled copies of a free-fermion system in arbitrary dimensions and perform post-selective Bell measurements on half of the corresponding sites across the two copies. Remarkably, the post-measurement state factorizes into a product of Bell pairs, establishing maximal interlayer entanglement entirely independent of the initial Gaussian state. 
 We derive this post-measurement state exactly for general particle-number-conserving fermionic Gaussian states, establishing both the validity and universality of the mechanism, with numerical simulations serving as consistency checks. This phenomenon arises from a robust interplay between fermionic statistics and Gaussianity, revealing a distinct fermionic route to measurement-induced maximal entanglement.
\end{abstract}
\maketitle

\section{Introduction}

Quantum entanglement is known to be intrinsic to quantum mechanics and has no classical analog, being famously associated with its ``spooky action at a distance''. Beyond its role in foundational thought experiments, a central question -- both conceptually and technologically -- is how to generate nonlocal entanglement between distant systems that have never interacted directly. This question is especially important because entanglement is a key resource for quantum communication, computation, and related quantum-information tasks~\cite{Nielsen_Chuang_2010, Horodecki_RMP_entanglement_2009, RMP_quantum_resource_2019}.

A particularly striking mechanism for generating such remote entanglement is entanglement swapping: by performing a joint Bell measurement on two intermediate systems that are each entangled with distant partners, one can effectively reconfigure the entanglement structure and transfer entanglement from local pairs to remote ones~\cite{bennett_teleport_1993, zukowski_swap_1993, bose_swap_1998, mattle_teleport_1996, bouwmmeester_teleport_1997, boschi_teleport_1998, furusawa_teleport_1998, pan_swap_1998}. This process underlies a broad range of quantum technologies, including quantum cryptography~\cite{gisin2002_cryptography} and quantum communication protocols such as quantum repeaters~\cite{azuma2023_repeater}.

More broadly, measurement itself plays a fundamental role among non-unitary processes in quantum mechanics. Beyond its essential function in extracting information from quantum systems \cite{kraus_1983, vogel_tomo_1989, smithey_tomo_1993, mlynek_measure_1997, white_tomo_1999, aharonov_weak_measure_1988, bamber_measure_2011},
measurement can also profoundly reshape quantum states and induce rich entanglement structures. In recent years, this perspective has motivated extensive studies of measurement-driven many-body dynamics, including measurement-induced entanglement transitions \cite{skinner_MIPT_2019, li_MIPT_2018, li_MIPT_2019, chan_projective_2019, 
2020_Altman,
Jian_2020,
2022_Pixley,Buchhold_2021},
critical states under measurements or dissipation \cite{Ashida_2016,Minoguchi_2022,
Yamamoto_2022,Garratt_2023,
alicea_measurement_altered_2023, 
Lee_2023,
jsk_measure_critical_2023,Weinstein_2023,
Zou_2023,
jian_measure_critical_2023,
liu_single_measure_gapless_2025,
zhang_critical_mix_measure_2025, Hoshino_stabilizer_cft_2025,
Tang_measure_critical_2024}, 
measurement induced entanglement in many-body systems\cite{potter_measure_induce_info_2024,Lin_2023,
Zhang_2024,
Negari_2024,
khanna2025}, 
and 
long-range entangled state preparation
\cite{
Lu_2022,
2023_Nat,
Tantivasadakarn_2024,
verresen2022Cat, 2024_non_Abelian,lee_2022decoding,Zhu_2023,Smith_2024}. Among the various measurement protocols explored in this context, Bell measurements are especially notable due to their central role in enabling quantum teleportation and entanglement swapping~\cite{bennett_teleport_1993, zukowski_swap_1993, bose_swap_1998, mattle_teleport_1996, bouwmmeester_teleport_1997, boschi_teleport_1998, furusawa_teleport_1998, pan_swap_1998,Antonini_2022,Milekhin_2024,Antonini_2023}.

Despite their fundamental and practical importance, most previous works on entanglement swapping have focused on few-body systems. Only recently have entanglement-swapping protocols been explored in many-body spin (qubit) systems~\cite{alicea_critical_teleport_2024, potter_measure_induce_info_2024, ashida_measurement_bcft_2024, 
Hoshino_stabilizer_cft_2025, 2025_Oshikawa_Bell, Huhtanen2025}. When the pre-measurement state is the ground state of a critical spin chain, the effects of imperfect teleportation and entanglement swapping have been analyzed~\cite{alicea_critical_teleport_2024, ashida_measurement_bcft_2024}, revealing universal scaling behaviors of correlations and entanglement that admit an interpretation in terms of boundary conformal field theory \cite{Cardy_bcft_2004}.

In contrast, the study of entanglement swapping in fermionic many-body systems remains far less developed. The intrinsic fermionic statistics, together with the nonlocal structure introduced by the Jordan–Wigner transformation, significantly complicate both analytical and numerical approaches. To the best of our knowledge, universal entanglement features arising from entanglement swapping in many-body fermionic systems have not been reported to date.

\begin{figure*}
\centering
\includegraphics[width=6.6in]{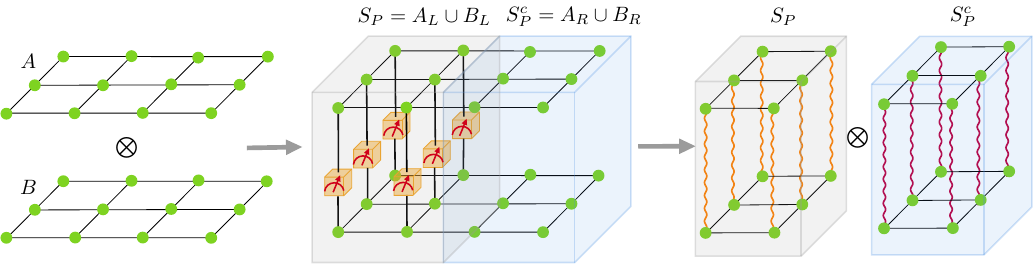}
\caption{Left: Two identical free-fermion layers $A$ (upper) and $B$(lower) in $d$ dimensions (illustrated for $d=2$), each with an even number of sites $L$. Each layer is prepared in a number-conserving Gaussian state at half filling. 
 Middle: Projective Bell measurements are performed on the subsystem $S_P$.
While $S_P$ can be chosen arbitrarily, here we take it to be the left half of the system, i.e., $S_P = A_L \cup B_L$,
where we denote the left (right) half of $A$ as $A_L$ ($A_R$), and similarly for $B$.
 For every site $i \in \{1,\dots,L\}$ in layer $A$, there is a dual site $\bar{i}$ in layer $B$. On each rung $(i,\bar{i}) \subset S_P$ we apply the Bell projector defined in Eq.~\eqref{Bell_rung}. 
 That is, each rung in $S_p$ is projected to the Bell state
   $\ket{+}_{i\bar i}
  = \tfrac{1}{\sqrt{2}}\big(c_i^\dagger + c_{\bar{i}}^\dagger\big)\ket{\mathrm{vac}}$.
Right: After $L/2$ such rung measurements, the post-selected state factorizes across the bipartition as a tensor product of a state on $S_P$ 
and one on its complement
$S_P^c=A_R\cup B_R$. On $S_P$ each rung $(i,\bar i)$ is in the Bell state $\ket{+}_{i\bar i}$ (orange lines), while on the unmeasured complement $S_P^c$ each rung $(i,\bar{i})$ is in the Bell state
  $\ket{-}_{i\bar i}
  = \tfrac{1}{\sqrt{2}}\big(c_i^\dagger - c_{\bar{i}}^\dagger\big)\ket{\mathrm{vac}}$ (purple lines).
}
\label{fig:sketch}
\end{figure*}

In the present work, we focus on conditional entanglement swapping between a fermionic Gaussian state and its identical copy, and analyze how the entanglement structure is reshaped when Bell measurements are post-selected on a uniform outcome, i.e., when each measured pair is projected onto the same Bell state. In contrast, standard (unconditional) entanglement swapping, which averages over all measurement outcomes according to the Born rule, constitutes a local-operations-and-classical-communication (LOCC) protocol and can only redistribute pre-existing entanglement rather than create new entanglement~\cite{Vedral_1997, Vidal_1998, Horodecki_entanglement_limit_2000}. By introducing post-selection, we explicitly break the LOCC structure, thereby allowing the total entanglement to increase under conditional entanglement swapping.

While it has been observed that conditional entanglement swapping in critical spin chains leads to universal logarithmic scaling of entanglement entropy~\cite{2025_Oshikawa_Bell}, no entanglement enhancement was found. Here, instead, we establish an analytical, state-independent result for particle-number-conserving fermionic Gaussian states in arbitrary spatial dimensions: 
As shown in Fig.~\ref{fig:sketch}, we consider a bilayer system composed of two identical, initially decoupled copies of a particle-number-conserving fermionic Gaussian state at half filling.
Projective Bell measurements are performed on half of the interlayer site pairs (``rungs''), which we take for concreteness to be those in the left half of the system. Upon post-selecting a uniform Bell-measurement outcome, the post-measurement state factorizes between the measured and unmeasured regions, while the unmeasured degrees of freedom are driven into a tensor product of Bell pairs across the two layers.
As a result, the unmeasured subsystem exhibits universally maximal interlayer entanglement.
This phenomenon is enforced solely by fermionic anti-commutation relations together with Gaussianity, rendering it entirely independent of the microscopic details of the initial state.

The rest of this work is organized as follows. In Sec.~\ref{Sec:MainResult}, we introduce the setup and summarize the main result of this work. In Sec.~\ref{Sec:Proof}, we provide a detailed proof of the universal phenomenon established in Sec.~\ref{Sec:MainResult}. Section~\ref{Sec:Numerics} presents numerical verifications of our analytical results and further explores universal features of the entanglement entropy when fewer than half of the sites are measured. In Sec.~\ref{Sec:AmpEHRelation}, we investigate the relationship between the probability of obtaining the post-measurement state and the entanglement properties of the initial state. Then we conclude in Sec.~\ref{Sec:Conclusion} including a discussion of how the post-selected wavefunction may be realized deterministically and possible generalizations of our setup.
There are also several appendices providing additional results and technical details. Appendix~\ref{app:plucker} presents an alternative proof of our main theorem that is potentially helpful for establishing a geometric description of the maximal entanglement swapping. Further details of our measurement protocol are given in Appendices~\ref{app:non_separable}, \ref{app:spin}, and \ref{appendix:OtherBell}. Appendix~\ref{app:imperfect} analyzes the case in which the second layer is an imperfect copy of the first layer. Appendix~\ref{app:amplitude} examines how the probability of obtaining the post-selected state depends on the properties of the initial state.

\section{Free fermions subjected to Bell measurements}
\label{Sec:MainResult}

\subsection{Setup and main result}\label{sec:setup}

We consider a bilayer setup in which each layer is a $d$-dimensional lattice with $L$ sites (see Fig.~\ref{fig:sketch}). For every site $i\in [L]\equiv\{1,\dots,L\}$ in the upper layer, there is a corresponding “dual” site $\overline{i}$ in the lower layer. The upper layer hosts a number-conserving free fermionic state $\ket{\Psi_0}$ with $N$ spinless fermions, and the lower layer hosts an \emph{identical copy} $\ket{\overline{\Psi}_0}=\ket{\Psi_0}$ obtained by relabeling $i\mapsto \overline{i}$. The two layers are initially decoupled, hence the total state is the tensor product
\begin{equation}  \ket{\Psi}_{\mathrm{double}}=\ket{\Psi_0}\otimes\ket{\overline{\Psi}_0}\, .
\end{equation}

\medskip
We then perform Bell measurements on each rung $(i,\bar{i})$ within the region $S_P$ (see Fig.~\ref{fig:sketch}). \footnote{For notational simplicity, we will hereafter use $i \in A_L$ to denote the rung $(i,\bar{i})$ itself in $S_P = A_L \cup B_L$, whenever no confusion arises. 
} 
Here we define the set of measured rungs as $S_P \coloneqq A_L \cup B_L = \{ i\overline{i} \}$, where $i \in A_L \subset [L]$ and $\bar i \in B_L \subset [L]$. $A_L$ and $B_L$ are the sets of measured sites on the upper and lower layers, respectively.

For each rung $(i,\overline{i})$, we define the Bell state
\be
\label{Bell_+}
  \ket{+}_{i\overline{i}}
  =\frac{1}{\sqrt 2}\big(c_i^\dag+c_{\overline i}^\dag\big)|\text{vac}\rangle
  =\frac{1}{\sqrt{2}}\big(\ket{01}_{i\overline{i}}+\ket{10}_{i\overline{i}}\big),
\ee
where the fermionic operators obey the canonical anticommutation relations
$\{c_i,c_j^\dagger\}=\delta_{ij}$ and $\{c_i,c_j\}=\{c_i^\dagger,c_j^\dagger\}=0$, and $\ket{\mathrm{vac}}$ is the Fock vacuum.
The associated projector is
\begin{equation}
\label{Bell_rung}
  P^{+}_{i\overline{i}}
  = \ket{+}_{i\overline{i}}\,{}_{i\overline{i}}\!\bra{+}.
\end{equation}
In terms of fermionic operators, we have
\begin{equation}
\label{Fermion_Berry}
P_{i\overline i}^+=\frac{1}{2}
\Big[
n_i(1-n_{\overline i})+n_{\overline i}(1-n_i)+c_i^\dag c_{\overline i}+c_{\overline i}^\dag c_i
\Big],
\end{equation}
with $n_i=c_i^\dagger c_i$.
The (unnormalized) post-measurement state for the outcome in which every measured rung is projected onto $\ket{+}$ is
\begin{equation}
  \prod_{i\in A_L} P^{+}_{i\overline{i}}\ket{\Psi}_{\mathrm{double}}.
\end{equation}
After renormalization, the post-selected state is

\begin{equation}
  \ket{\Psi_{\mathrm{post}}}
  = \frac{\displaystyle \prod_{i\in A_L} P^{+}_{i\overline{i}}\ket{\Psi}_{\mathrm{double}}}
         {\displaystyle \Big\|\prod_{i\in A_L} P^{+}_{i\overline{i}}\ket{\Psi}_{\mathrm{double}}\Big\|}\,.
\end{equation}
The measured part factorizes from the rest as follows
\begin{equation}
\label{eq:post_state}
  \ket{\Psi_{\mathrm{post}}}
  = \ket{\Psi_{\mathbf m,+}}\otimes \ket{\Psi_{\mathrm{remain}}},
  \quad
  \ket{\Psi_{\mathbf m,+}}=\bigotimes_{i\in A_L}\ket{+}_{i\overline{i}},
\end{equation}
up to an overall phase, and $\ket{\Psi_{\mathrm{remain}}}$ is normalized and supported on the unmeasured region $A_R = [L]/A_L$.  
This form holds provided the post-selection amplitude
$\big\|\big(\prod_{i\in A_L} P^{+}_{i\overline{i}}\big)\ket{\Psi}_{\mathrm{double}}\big\|$ is nonzero; otherwise the outcome has zero probability and no post-selected state is defined.

Since the Bell measurements effectively couple the two layers, the protocol generically generates inter-layer entanglement. Remarkably, in our setting the post-selected state exhibits a rigid and universal entanglement pattern, captured by the following theorem:
\begin{theorem}
\label{corollary:post_measure}
	Let $\ket{\Psi_0}$ be a (number-conserving) free spinless fermion state on $L$ sites with particle number $N\in\{0,\dots,L\}$. Consider the doubled state $\ket{\Psi_{\mathrm{double}}}=\ket{\Psi_0}\otimes\ket{\overline{\Psi}_0}$ on a bilayer with dual sites $(i,\overline{i})$, where $\ket{\overline{\Psi}_0}$ is an identical copy of $|\Psi_0\rangle$. 
Postselect on the Bell outcome $\ket{+}_{i\overline{i}}$ on half of the rungs $\{i\overline{i}\}$, with $i\in A_L\subset [L]$, $|A_L|=L/2$.
Then, up to an overall phase, the unmeasured complement $A_R=[L]\setminus A_L$ collapses to
	\[
	|\Psi_{\rm{remain}}\rangle = \begin{cases}
		\otimes_{i \in A_R}  |-\rangle_{i\bar i}  , & N = \frac{L}{2}, \\ 
		0 , & N \neq \frac{L}{2},
	\end{cases}
	\]
	i.e. the full postselected state is
	\[
	| \Psi_{\rm post} \rangle = \begin{cases}
		\left( \otimes_{i \in A_L} |+\rangle_{i\bar i} \right) \otimes \left( \otimes_{i \in A_R} |-\rangle_{i\bar i} \right) , & N = \frac{L}{2}, \\ 0 , & N \neq \frac{L}{2}.
	\end{cases}
	\]
Here $|\pm\rangle_{i\bar i} = \frac{1}{\sqrt{2}} \left( |01\rangle_{i \overline{i}} \pm |10\rangle_{i \overline{i}} \right)$ are the local Bell states defined on a rung $\{i\bar i\}$.  
When $| \Psi_{\rm post} \rangle=0$, it means that the corresponding postselection has zero probability of occurrence.
\end{theorem}

In other words, for any ``nontrivial'' number-conserving free-fermion state $\ket{\Psi_0}$ at half filling $N=L/2$, if we postselect the Bell outcome $\bigotimes_{i\in A_L}\ket{+}_{i\overline{i}}$ on half of the rungs ($|A_L|=L/2$), then the unmeasured complement collapses to a product of Bell pairs,
\[
\ket{\Psi_{\mathrm{remain}}}
=\bigotimes_{i\in A_R}\ket{-}_{i\overline{i}},
\]
up to an overall phase. Here, ``non-trivial'' means that the initial state $|\Psi_0\rangle$ hosts non-zero amplitude for  projecting onto both of the computational basis $\prod_{i \in A_L} c_i^\dagger | {\rm{vac}} \rangle_{\rm{upper}}$ and $\prod_{i \in A_R} c_i^\dagger | {\rm{vac}} \rangle_{\rm{upper}}$. Further details and explicit conditions will be discussed later.

\subsection{Wavefunction of doubled state in a coupled basis of two layers}

\subsubsection{The wavefunction in computational basis}

We work with number-conserving (U(1)-symmetric) fermionic Gaussian states, i.e., Slater determinants. 
Any pure state on $L$ spinless orbitals with $N$ particles can be written as
\begin{equation}
	|\Psi_0\rangle = \prod_{q=1}^{N} b_q^\dagger \vac . 
\end{equation}
Here $\vac$ is the vacuum state
and the creation operators
\begin{equation}
  b_q^\dagger=\sum_{i=1}^L \Phi_{qi}\, c_i^\dagger
\end{equation}
define orthonormal single-particle orbitals. In vector notation,
\begin{equation}
  \mathbf b^\dagger=\Phi\,\mathbf c^\dagger, \quad
  \mathbf c^\dagger=(c_1^\dagger,\dots,c_L^\dagger)^{\mathsf T},\ 
  \mathbf b^\dagger=(b_1^\dagger,\dots,b_L^\dagger)^{\mathsf T},
\end{equation}
where $\Phi \in U(L)$ is a unitary matrix whose entries $\Phi_{qi}$ represent the single-particle wavefunctions, i.e., the amplitude of orbital $q$ on site $i$.
In the standard computational (real-space) basis, the single-layer initial state $|\Psi_0\rangle$ has the following representation
\begin{equation}
	\begin{aligned}
		| \Psi_0 \rangle & = \sum_{[i_1, \cdots, i_N] \subset [L]} \psi_{[i_1, \cdots, i_N]} c_{i_1}^\dagger \cdots c_{i_N}^\dagger \vac_{\rm upper} 
		\\ & = \sum_{A \subset [L], |A|=N} \Delta_A \prod_{i \in A} c_i^\dagger  \vac_{\rm upper} .
	\end{aligned}
\end{equation}
Here $\Delta_A = \det \Phi([1, \cdots, N], [i_1, \cdots, i_N])$ is the \textit{Slater determinant} for a given set of occupied sites $A = [i_1, \cdots, i_N] $,
i.e., it corresponds to the $N\times N$
minor of $\Phi$ with row set $[N]=\{1,\cdots,N\}$ and column set $A$,
and $\vac_{\rm upper}$ represents the vacuum state defined on the upper layer. 

For the doubled state $| \Psi \rangle_{\rm double} = | \Psi_0 \rangle \otimes | \overline{\Psi}_0 \rangle$ 
(two identical copies, hence $2N$ particles), it is convenient to write
\begin{equation}
  \begin{pmatrix}\mathbf b^\dagger \\ \overline{\mathbf b}^\dagger\end{pmatrix}
  = \mathrm{diag}(\Phi,\Phi)
    \begin{pmatrix}\mathbf c^\dagger \\ \overline{\mathbf c}^\dagger\end{pmatrix},
  \qquad
  \overline{\mathbf c}^\dagger=(c_{\overline{1}}^\dagger,\dots,c_{\overline{L}}^\dagger)^{\mathsf T},
\end{equation}
which gives
\begin{equation}\label{eq:rep_double_psi_1}
		| \Psi \rangle_{\rm double} =
		\sum_{ \substack{A \subset [L], |A| = N \\ B \subset [L], |B| = N} } \Delta_A \Delta_B \prod_{i \in A} c_{i}^\dagger \prod_{\overline{j} \in B} c_{\overline{j}}^\dagger \vac 
\end{equation}
where $\vac = \vac_{\rm upper} \otimes \vac_{\rm lower}$ is the bilayer vacuum as a product of the vacua in the upper and lower layers.

\subsubsection{The $SU(2)$ rotation that couples the two layers}

To prove Theorem~\ref{corollary:post_measure}, it is convenient to introduce, on each rung $(i,\overline{i})$, the composite creation operators
\begin{equation}
\widetilde{w}_{i,+}^\dagger = \alpha c_i^\dagger + \beta c_{\overline{i}}^\dagger \,,
\quad
\widetilde{w}_{i,-}^\dagger = -\beta^* c_i^\dagger + \alpha^* c_{\overline{i}}^\dagger 
\end{equation}
with $|\alpha|^2+|\beta|^2=1$, so that the transformation is an $SU(2)$ rotation. These composite operators generate the orthonormal local basis
\begin{equation}\label{eq:def_w_basis}
\begin{aligned}
|b_0\rangle & = | 00 \rangle_{i \overline{i}}, \quad
|b_1\rangle = | \widetilde{+} \rangle_{i\bar i} = \widetilde{w}_{i,+}^\dagger | 00 \rangle_{i \overline{i}}, \\
|b_2\rangle & = | \widetilde{-} \rangle_{i\bar i} = \widetilde{w}_{i,-}^\dagger | 00 \rangle_{i \overline{i}}, \\
|b_3\rangle & = | 11 \rangle_{i\bar i} = \widetilde{w}_{i,+}^\dagger \widetilde{w}_{i,-}^\dagger | 00 \rangle_{i \overline{i}}, 
\end{aligned}
\end{equation}
which forms a complete basis for the two-site Hilbert space on rung $(i,\overline{i})$. 
In the special case $\alpha = \beta = 1/\sqrt{2}$, the composite operators $\widetilde{w}_{i,\pm}^\dagger$ reduce, up to an overall phase, to the creation operators of the local Bell states $|+\rangle_{i\overline{i}}$ and $|-\rangle_{i\overline{i}}$, respectively.

Applying the local basis rotation to the full bilayer, the single-particle basis becomes
\begin{equation}
	\widetilde{M}_L 
	\left( \begin{matrix} 
		\widetilde{\mathbf{w}}_{+}^\dagger \\ \widetilde{\mathbf{w}}_{-}^\dagger 
	\end{matrix} \right) 
	= 
	\left( \begin{matrix}
		\mathbf{c}^\dagger \\ \overline{\mathbf{c}}^\dagger
	\end{matrix} \right) \,, 
	\ 
	\widetilde{M}_L = \left(\begin{matrix}
		\alpha\mathbf{1}_{L\times L} & \beta\mathbf{1}_{L\times L} \\ -\beta^*\mathbf{1}_{L\times L} & \alpha^*\mathbf{1}_{L\times L}
	\end{matrix}\right) ,
    \nonumber
\end{equation}
where $\mathbf{1}_{L \times L}$ denotes the $(L \times L)$ identity matrix. 
Based on this expression, we can write down the sub-orbital matrix restricted to the occupied band as
\begin{equation}
\begin{aligned}
\left(\begin{matrix}
	\Phi_{[N], [L]} & \mathbf{0} \\ \mathbf{0} & \Phi_{[N], [L]}
\end{matrix}\right) 
\widetilde{M}_L 
= \left(\begin{matrix} \alpha \Phi_{[N], [L]} & \beta \Phi_{[N], [L]} \\ -\beta^*\Phi_{[N], [L]} & \alpha^* \Phi_{[N], [L]} \end{matrix}\right) . 
\end{aligned}
\nonumber
\end{equation}
Due to the particle-number conservation, only product states of the form
\begin{equation}
| \widetilde{\mathcal{B}}_{Q_+, Q_-} \rangle = \prod_{i \in Q_+} \widetilde{w}_{i, +}^\dagger \prod_{i \in Q_-} \widetilde{w}_{i, -}^\dagger | \text{vac} \rangle 
\end{equation}
with $Q_{\pm}\subset [L]$ and $| Q_+ | + | Q_-| = 2N$ (note that $Q_+\cap Q_-\neq\emptyset$ is allowed), 
can appear with nonzero amplitude. For such basis, the many-body coefficient of wavefunction is given by the following Slater determinant
\begin{equation}\label{eq:bell_basis_slater}
\begin{aligned}
\psi_{\mathcal{B}} & = \det \left(\begin{matrix} \alpha \Phi_{[N], [L]} & \beta \Phi_{[N], [L]} \\ -\beta^*\Phi_{[N], [L]} & \alpha^* \Phi_{[N], [L]} \end{matrix}\right)
\\ & = \det (\widetilde{M}_N) \det \left(\begin{matrix}
\Phi_{[N], Q_+} & \mathbf{0} \\ \mathbf{0} & \Phi_{[N], Q_-}
\end{matrix}\right) 
\\ & = \det \left(\begin{matrix}
	\Phi_{[N], Q_+} & \mathbf{0} \\ \mathbf{0} & \Phi_{[N], Q_-}
\end{matrix}\right) 
\end{aligned}
\end{equation}
where we have used $\det(AB) = \det (A) \det (B)$ for square matrices $A$ and $B$, and $\det (\widetilde{M}_N)=1$ since $\widetilde{M}_N$ is basically a product of local $SU(2)$ rotations. In the specific case that $\alpha = \beta = \frac{1}{\sqrt{2}}$, i.e., the local two-site basis is rotated to the Bell basis $|+\rangle_{i\overline{i}}$ and $|-\rangle_{i\overline{i}}$, this representation gives the amplitude of the doubled state when the left half is projected onto $\bigotimes_{i\in A_L}\ket{+}_{i\bar i}$.

\section{Proof of Theorem~\ref{corollary:post_measure}: universal post-measurement state}
\label{Sec:Proof}

As seen in Eq.\eqref{eq:bell_basis_slater}, the amplitude of $SU(2)$ rotated basis is the determinant of a block-diagonal matrix, which makes the structure of the doubled state $\ket{\Psi}_{\mathrm{double}}=\ket{\Psi_0}\otimes\ket{\overline{\Psi}_0}$ transparent. Although the main focus of this work is on Bell-state measurement, the universal structure of the $SU(2)$ rotation motivates us to propose the following more general statement, for which Theorem~\ref{corollary:post_measure} is a direct corollary.

\begin{theorem}\label{theorem:2}
Let $\ket{\Psi_0}$ be a (number-conserving) free spinless fermionic state on a $d$-dimensional lattice with sites $i=1,\dots,L$, and with fixed particle number $N \in {0,\dots,L}$. Consider the doubled state $| \Psi \rangle_{\rm double} = | \Psi_0 \rangle \otimes | \overline{\Psi}_0 \rangle$, where $|\overline{\Psi}_0\rangle = |\Psi_0\rangle$ is an
identical copy living on the dual sites $\overline{i}$. For each rung $(i,\overline{i})$,
we use the orthonormal two-site basis obtained by the $SU(2)$ rotation 
defined in Eq.~\eqref{eq:def_w_basis}.  
	
Fix a subset $A_L \subset [L] = \{ 1, \dots, L \}$ with $|A_L| = L/2$ (hence $L$ is even), and denote its complementary as $A_R = [L] / A_L$. Define the product state on the measured rungs
	\[
	|\widetilde{\Psi}_{\mathbf{m}, +}\rangle := \otimes_{i \in A_L}|\widetilde{+}\rangle_{i\bar i} .
	\]
	Then 
	\begin{enumerate}
		\item Among all product states of the orthogonal local two-site basis $\left\{ |00\rangle_{i\overline{i}}, \  |11\rangle_{i\overline{i}}, \ |\widetilde{+}\rangle_{i\bar i}, \ |\widetilde{-}\rangle_{i\bar i} \right\}$ 
        that agree with $\ket{\widetilde{\Psi}_{\mathbf m,+}}$ on the measured rungs
$S_P = \{(i,\overline{i}) \mid i \in A_L\}$, 
there is a unique product state with nonvanishing amplitude, namely
		\[
		|\widetilde{\mathcal{B}}_{\text{dual}}\rangle = \left[ \otimes_{i \in A_L} |\widetilde{+}\rangle_{i\bar i} \right] \otimes \left[ \otimes_{i \in A_R} |\widetilde{-}\rangle_{i\bar i} \right] ,
		\]
        where $A_R = [L]/A_L$ is the complementary of $A_L$. 
		All other product states are orthogonal to $|\widetilde{\mathcal{B}}_{\text{dual}}\rangle$ and have zero amplitude. 
		\item The product state $|\widetilde{\mathcal{B}}_{\text{dual}}\rangle$ hosts a non-vanishing amplitude, if and only if both of the following computational-basis states have non-zero amplitude in $|\Psi_0\rangle$: 
			\[
			\prod_{i \in A_L} c_i^\dagger | {\rm{vac}} \rangle_{\rm{upper}}, 
			\quad 
			\prod_{i \in A_R} c_i^\dagger | {\rm{vac}} \rangle_{\rm{upper}} 
			\]
  where $\ket{\mathrm{vac}}_{\mathrm{upper}}$ is the vacuum state of the (original) upper layer.
	\end{enumerate}
\end{theorem}

The above theorem can be easily proven by using the relation of the wavefunction under the $SU(2)$ rotation discussed in the previous section. In particular, let us begin by introducing the following lemma. 
\begin{lemma}\label{lemma:1_det}
The determinant in Eq.~\eqref{eq:bell_basis_slater} is nonzero if and only if $|Q_+| = |Q_-| = N$. 
\end{lemma}
\noindent \textit{Proof. } 
Recall that $|Q_+|+|Q_-|=2N$. Write
\[
\mathcal M \;=\;
\begin{pmatrix}
  \Phi_{[N],Q_+} & \mathbf 0 \\
  \mathbf 0       & \Phi_{[N],Q_-}
\end{pmatrix},
\]
which is a $2N\times 2N$ matrix assembled from the two $N\times |Q_\pm|$ blocks.
If $|Q_+|>N$ (hence $|Q_-|<N$), the $N\times |Q_+|$ block $\Phi_{[N],Q_+}$ has more columns than rows, so its columns are linearly dependent. Because $\mathcal M$ is block–diagonal, any linear dependence among the columns of $\Phi_{[N],Q_+}$ extends to a linear dependence among the columns of $\mathcal M$, implying $\det \mathcal M=0$. The same argument applies if $|Q_-|>N$. Therefore, a necessary condition for $\det\mathcal M\neq 0$ is $|Q_+|=|Q_-|=N$.
Conversely, when $|Q_+|=|Q_-|=N$, the matrix $\mathcal M$ is block–diagonal with two $N\times N$ square blocks, hence
\[
\det\mathcal M=\det\!\big(\Phi_{[N],Q_+}\big)\,\det\!\big(\Phi_{[N],Q_-}\big),
\]
which is nonzero precisely when both minors are nonzero. This proves the claim.
\qed

\medskip
Although this property is mathematically elementary, it imposes a strong constraint, as stated in the following proposition.

\begin{proposition}
In the doubled initial state $|\Psi_{\rm double}\rangle = |\Psi_0\rangle \otimes |\overline{\Psi}_0\rangle$, among all product basis states $| \widetilde{\mathcal{B}}_{Q_+, Q_-} \rangle = \prod_{i \in Q_+} \widetilde{w}_{i, +}^\dagger \prod_{i \in Q_-} \widetilde{w}_{i, -}^\dagger | \text{vac} \rangle$ defined by the orthogonal local two-site basis $|b_k\rangle, k = 1,2,3,4$ in Eq.~\eqref{eq:def_w_basis}, only those satisfy $|Q_+| = |Q_-| = N$ have non-vanishing wavefunction. Here $N$ is the occupied particle number in a single copy of the initial state, $|\Psi_0\rangle$.
\end{proposition}
If the initial state is at half-filling, this constraint immediately leads to Statement~1 of Theorem~\ref{theorem:2}. To complete the proof, we further demonstrate that only the half-filling case can host a non-vanishing wavefunction upon projection onto the product of Bell pairs $|\widetilde{\Psi}_{\mathbf{m}, +}\rangle = \otimes_{i \in A_L}|\widetilde{+}\rangle_{i}$. To show this, we focus on those basis states that can have non-vanishing amplitudes. They must satisfy the following condition
\begin{equation}\label{eq:require_occupation}
A_L \subset Q_+ , \quad A_L \cap Q_- = \emptyset, \quad | Q_+ | + | Q_-| = 2N,
\end{equation}
i.e., all pairs on sites $\{i, \overline{i}\}, i \in A_L$ ($A_L\subset [L], |A_L|=L/2$) are in the state of $|\Psi_{\mathbf{m}, +}\rangle = \otimes_{i \in A_L}|\widetilde{+}\rangle_{i}$. 
In particular, under the above condition, we have the following lemma.

\begin{lemma}\label{lemma:2_rotation}
Under the condition in Eq.~\eqref{eq:require_occupation}, the determinant in Eq.~\eqref{eq:bell_basis_slater} vanishes whenever the system is not at half filling, i.e., for $N \neq L/2$.
\end{lemma}
\noindent \textit{Proof. } The requirement of $A_L \subset Q_+$ indicates $|A_L| = L/2 \le |Q_+|$. For $N<L/2$ (fillings below half-filling), $|Q_+| \neq |Q_-|$, so the determinant vanishes by Lemma~\ref{lemma:1_det}. For $N > L/2$ (filling above half-filling), there are two cases to discuss. First, when $|Q_+| \neq N$, the determinant in Eq.~\eqref{eq:bell_basis_slater} vanishes by Lemma~\ref{lemma:1_det}, just as in the case of $N<L/2$. Second, when $|Q_+|=N$, the determinant is already in block form with four $(N \times N)$ blocks. However, since $|A_L|=L/2<N=|Q_-|$ and $A_L,Q_-\subseteq [L]$, we have $|A_L|+|Q_-|>L$, so the condition $A_L \cap Q_- = \emptyset$ can not be satisfied and this case is forbidden, i.e., the corresponding wavefunction must vanish.
$\qed$

\begin{lemma}\label{lemma:3_rotation}
In the half-filled case $N=L/2$, the amplitude in Eq.~\eqref{eq:bell_basis_slater} reduces to
\[
  \psi_{\mathcal B}=\,\det\!\big(\Phi_{[N],A_L}\big)\,\det\!\big(\Phi_{[N],A_R}\big),
\]
where $A_R=[L]\setminus A_L$ is the complementary of $A_L$.
\end{lemma}

\noindent\textit{Proof.}
When $N=L/2$, the constraints~\eqref{eq:require_occupation} uniquely determine
$Q_+=A_L$ and $Q_-=A_R$. Substituting these into Eq.~\eqref{eq:bell_basis_slater} yields the stated expression. \qed

\medskip

\noindent\textbf{Proof of Theorem~\ref{theorem:2}.}
By Lemma~\ref{lemma:3_rotation}, at half filling the relevant amplitude is
$\det(\Phi_{[N],A_L})\,\det(\Phi_{[N],A_R})$.
The condition $\det(\Phi_{[N],A_L})\,\det(\Phi_{[N],A_R})\neq 0$ is
equivalent to requiring that both computational–basis configurations
$\big(\prod_{i\in A_L} c_i^\dagger\big)\ket{\mathrm{vac}}_{\mathrm{upper}}$ and
$\big(\prod_{i\in A_R} c_i^\dagger\big)\ket{\mathrm{vac}}_{\mathrm{upper}}$
appear with nonzero coefficients in $\ket{\Psi_0}$.
Combined with Lemma~\ref{lemma:2_rotation} (which rules out $N\neq L/2$),
this proves the theorem. \qed

\medskip

\noindent\textbf{Proof of Theorem~\ref{corollary:post_measure}.}
Theorem~\ref{theorem:2} reveals the universal structure of the post-measurement state after projecting onto a coupled basis of the two layers that obtained by a $SU(2)$ rotation. As we have mentioned, when the $SU(2)$ rotation is chosen with parameters $\alpha=\beta=\frac{1}{\sqrt{2}}$, the protocol in Theorem~\ref{theorem:2} corresponds to postselecting the state to the product of Bell states: $\left( \otimes_{i \in A_L} |+\rangle_{i\bar i} \right) \otimes \left( \otimes_{i \in A_R} |-\rangle_{i\bar i} \right) $. This is equivalent to  performing the Bell measurements $P^+_{i\overline{i}}=\ket{+}_{i\overline{i}}\,{}_{i\overline{i}}\!\bra{+}$  over rungs $\{i\overline{i}\}$, where $i \in A_L$. In summary, Theorem~\ref{corollary:post_measure} is a direct corollary of Theorem~\ref{theorem:2}.\qed

\medskip
We now further clarify the conditions under which the initial doubled state should be regarded as ``trivial'', namely when the corresponding post-selected outcome occurs with zero probability. Although it is not straightforward to formulate a simple sufficient criterion that guarantees a given initial state $|\Psi_0\rangle$ is nontrivial, our numerical tests indicate that typical ground states of both critical and gapped free-fermion chains possess nonzero amplitudes for the two relevant computational-basis states and are therefore nontrivial in this sense.

Moreover, certain trivial states have vanishing probability amplitude of projecting onto $\prod_{i \in A_L} c_i^\dagger | {\rm{vac}} \rangle_{\rm{upper}}$ and $\prod_{i \in A_R} c_i^\dagger | {\rm{vac}} \rangle_{\rm{upper}}$. In particular, we find a necessary condition for $|\Psi_0\rangle$ to be non-trivial is that it is non-separable with respect to a general real-space bipartition, as formalized in the following proposition:

\begin{proposition}\label{proposition}
	If the half-filling state $|\Psi_0\rangle$ can be factorized to a simple product of two parts of the entire layer of even size $L$ (the filled particle number is $N=L/2$), i.e. the state can be written as
	\begin{equation}
		| \Psi_0 \rangle = | \Psi_G \rangle \otimes | \Psi_{G^c} \rangle, \quad G \cap G^c = \emptyset, \ G \cup G^c = [L] . 
	\end{equation}
	then, at least one of the probability amplitudes of projecting $|\Psi_0\rangle$ onto $\prod_{i \in A_L} c_i^\dagger | {\rm{vac}} \rangle_{\rm{upper}}$ or $\prod_{i \in A_R} c_i^\dagger | {\rm{vac}} \rangle_{\rm{upper}}$ vanishes for arbitrary choice of the site set $A_L \subset [L], A_R = [L]/A_L$ with $|A_L|=|A_R| = L/2$. 
    The only exception is the symmetric case where both $|\Psi_G\rangle$ and $|\Psi_G^c\rangle$ are half-filled, and the chosen measurement region splits each subset evenly, i.e. $|T| = {|G|}/{2}$ and $|T^c| =  {|G^c|}/{2}$. Here $T = G \cap A_L$ and $T^c = G^c \cap A_L$ are the sets of measured rungs in $G$ and $G_c$, respectively.
\end{proposition}
\noindent\textit{Proof. } Suppose $|G| = m, |G^c| = L-m = 2N-m, m \in [L] = \{1, \dots, L\}$, and the number of occupied spinless fermions in $G$ and $G^c$ are $k$ and $N - k$ respectively. Under this setting the single-layer orbital matrix of the initial state is 
\[
\begin{pmatrix}
\Phi^{\text{sub}}_{[k], G} & \mathbf{0} \\ \mathbf{0} & \Phi^{\text{sub}}_{[k+1, N], G^c} 
\end{pmatrix}
\]
Since the projection is applied onto $G$ and $G^c$ separately, the projections of the orbital matrix of the state $|\Psi_0\rangle$ to the target basis is given by
\[ 
\begin{pmatrix}
    \begin{pmatrix}
        \Phi^{\text{sub}}_{[k], T} & \mathbf{0}  \\ \mathbf{0} & \Phi^{\text{sub}}_{[k], G/T}
    \end{pmatrix} & \mathbf{0} \\
    \mathbf{0} & \begin{pmatrix}
	\Phi^{\text{sub}}_{[k+1, N], T^c} & \mathbf{0} \\ \mathbf{0} & \Phi^{\text{sub}}_{[k+1, N], G^c/T^c} 
    \end{pmatrix}
\end{pmatrix}
\]
Similar to the derivation in Lemma~\ref{lemma:2_rotation}, the determinants of these two (sub)matrices after projection can have nonzero value only when $k = |T| = |G/T|$ and $N-k = |T^c| = |G^c/T^c|$. From $k = |T| = |G/T| = {|G|}/{2}$ and $N-k = |T^c| = |G^c/T^c| = (2N-m)/2$ (hence $|G|=m$ and $|G^c|=2N-m$ are even), both $|\Psi_G\rangle$ and $|\Psi_G^c\rangle$ are at half-filling. 
This is the condition under which the probability amplitudes for projecting onto $\prod_{i \in A_L} c_i^\dagger | {\rm{vac}} \rangle_{\rm{upper}}$ and $\prod_{i \in A_R} c_i^\dagger | {\rm{vac}} \rangle_{\rm{upper}}$ are both non-zero. 
Otherwise, at least one of the determinants will have zero value, and by definition the probability amplitude of projecting onto the computational basis vanishes. 
$\qed$

Consequently, for a general spatially separable state, 
the corresponding post-measurement amplitude vanishes, i.e., $|\Psi_{\rm post}\rangle = 0$, 
unless the state satisfies the specific conditions stated in Proposition~\ref{proposition}.
It should be noticed that the converse is not true: non-separability does not guarantee a non-vanishing post-measurement state, see Appendix~\ref{app:non_separable} for a counterexample. More importantly, as we will show in Sec.\ref{Sec:AmpEHRelation}, we establish a relation between the post-measurement probability and the entanglement property of the initial single-layer state, which provides a more quantitative understanding of the non-triviality condition.

Finally, we would like to point out that the roles of Bell states $|+\rangle = \frac{1}{\sqrt{2}} ( |01\rangle + |10\rangle )$ and $|-\rangle = \frac{1}{\sqrt{2}} ( |01\rangle - |10\rangle )$ in Theorem~\ref{corollary:post_measure} are interchangeable.
This follows from the fact that the statement of Theorem~\ref{theorem:2} remains valid under a general local $SU(2)$ rotation. In particular, if one performs the local projection
\begin{equation}
P^-_{i \overline{i}} = | - \rangle_{i \overline{i}}\, {}_{i \overline{i}}\langle - | = \frac{1}{2} (| 01 \rangle_{i \overline{i}} - | 10 \rangle_{i \overline{i}}) ({}_{i \overline{i}}\langle 01 | - {}_{i \overline{i}}\langle 10 |)
\end{equation}
on exactly half of the rungs, with site indices $i \in A_L \subset [L]$ and $|A_L| = L/2$, then the post-measurement state takes the form
\begin{equation}
| \Psi_{\rm post} \rangle = \left[ \otimes_{i \in A_L} |-\rangle_{i\bar i} \right] \otimes \left[ \otimes_{i \in A_R} |+\rangle_{i\bar i} \right] 
\end{equation}
if and only if the initial state $|\Psi_0\rangle$ is half-filled, i.e., $N = L/2$. Here $A_R = [L]/A_L$ is the complementary of $A_L$. For $N \neq L/2$, the amplitude of the post-measurement state vanishes. 

Before concluding this section, we remark that Appendix~\ref{app:plucker} presents an alternative proof of Theorem~\ref{corollary:post_measure}, based on the Pl\"ucker relation on the Grassmannian.

\section{Numerical results based on the correlation matrix technique}
\label{Sec:Numerics}

In the above discussions, we have fully characterized the structure of the post-selected wavefunction when $L/2$ or more Bell measurements are performed. When the number of Bell measurements is fewer than $L/2$, however, the structure of the post-selected wavefunction becomes more intricate. In this regime, we will employ numerical simulations to analyze the resulting wavefunction and identify its universal features.

\medskip

Since the states are free (Gaussian), all many–body quantities can be reduced to
the two–point (Majorana) correlation matrix
\be
  \Gamma_{ij} \;=\; \langle a_i a_j\rangle - \delta_{ij},
  \quad \text{where } a_k^\dagger=a_k,\ \ \{a_i,a_j\}=2\delta_{ij},
\ee
via Wick’s theorem. This is the standard \textit{correlation matrix} framework
(see, e.g., Refs.~\cite{Peschel2003,bravyi2005}). To implement a projective Bell
measurement we postselect with the Gaussian projector
\(\rho_M=\prod_{i\in A_L} P^+_{i\overline{i}}\), which can be considered as a reduced density matrix, and define the post–measurement state as
\begin{equation}
  \rho_{\mathrm{post}}
  \;=\; \frac{\rho_M\,\rho_0\,\rho_M}{\Tr(\rho_M\,\rho_0\,\rho_M)}\,,
  \qquad
  \rho_0=\bigl(\ket{\Psi}\bra{\Psi}\bigr)_{\mathrm{double}} \, ,
\end{equation}
where the trace term provides the normalization.

Products of Gaussian density matrices remain Gaussian. If \(\Gamma_1\) and
\(\Gamma_2\) are the correlation matrices of two normalized Gaussian density
matrices \(\rho_1\) and \(\rho_2\), the correlation matrix of the (normalized)
product \(\widetilde{\rho}\propto \rho_1\rho_2\) is given by
the composition rule~\cite{Fagotti2010_disjoint}
\be
  \widetilde{\Gamma}
  \;=\; \Gamma_1 \times \Gamma_2
  \;=\; \mathbf{1}
        - ( \mathbf{1}-\Gamma_2 )
          \bigl( \mathbf{1} + \Gamma_1 \Gamma_2 \bigr)^{-1}
          ( \mathbf{1}-\Gamma_1 ) .
\ee
In our case one applies this rule twice to obtain
\(\Gamma_{\mathrm{post}} = (\Gamma_M \times \Gamma_0) \times \Gamma_M\),
where \(\Gamma_0\) and \(\Gamma_M\) are the correlation matrices of
\(\rho_0\) and \(\rho_M\), respectively.

\begin{figure}\centering
\includegraphics[width=\columnwidth]{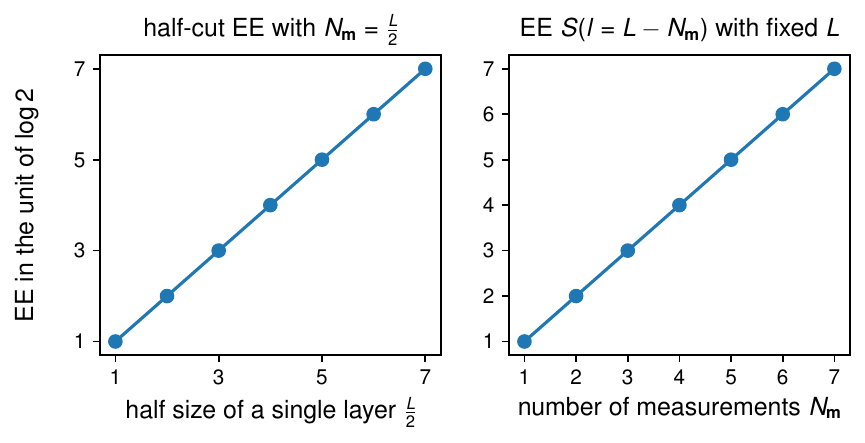}
\caption{Numerical results of the EE between the unmeasured part on a single layer and the rest, after performing uniform Bell measurements $P_i^+$ ($i \in A_L, |A_L| = N_{\mathbf{m}}$). Left: The measurements are performed precisely on half system with $N_{\mathbf{m}} = L/2$, with varying the total system size. The EE displays a linear dependence $S = \frac{L}{2} \log 2$ with the system size of a single layer $L$. Right: The number of performed measurements $N_{\mathbf{m}} \le L/2$, with a fixed system size $2L=28$. The EE displays a linear dependence $S = N_{\mathbf{m}} \log 2$ with the total system size of a single layer $L$. Here we consider the ground state to the critical Hamiltonian $H_0 = -\frac{1}{2} \sum_{i} c_i^\dagger c_{i+1}^\dagger + h.c.$ with open boundary conditions, and we have tested that other states, including gapped ground states, leads to the same behavior. The size of the total bilayer system is $2L$.}
\label{fig:EE_numeric_main}
\end{figure}

Using the composition rule above, we obtain the postselected Gaussian state and
compute the entanglement entropy (EE) for subsystem $A_R$ which corresponds to the unmeasured region in system $A$ (See Fig.\ref{fig:sketch}):
\be
  S_{A_R}\;=\;-\Tr\!\big(\rho_{\mathrm{post},A_R}\log \rho_{\mathrm{post},A_R}\big),
\ee
where $\rho_{\mathrm{post},A_R}=\Tr_{\bar A_R}\rho_{\mathrm{post}}$, and
$\bar A_R$ denotes the complement of $A_R$.

Consistent with Theorem~\ref{corollary:post_measure}, for a half-filled initial state and exactly $N_{\mathbf m}=L/2$ Bell projections on the rungs $(i,\overline{i})$ in $S_P$, our analysis of the correlation matrix $\Gamma$ confirms that the postselected state coincides with the analytically predicted product of local Bell pairs.
As shown in the left panel of Fig.~\ref{fig:EE_numeric_main}, performing $N_{\mathbf m}=L/2$ Bell measurements within the regions $S_P$ (see Fig.~\ref{fig:sketch}) yields an entanglement entropy for subsystem $A_R$ of
\be
\label{SA_half_L}
S_{A_R}(L/2)=\frac{L}{2}\log 2,
\ee
as expected, since each rung in the complementary region $S_P^c$ forms a Bell pair.
In contrast, when the initial filling is not  $L/2$, the postselection amplitude vanishes, again in agreement with Theorem~\ref{corollary:post_measure}.

We further explore the more general case with $N_{\mathbf m}< L/2$ Bell projections in half-filled states, where only a fraction of the rungs (less than one half) are measured. Although the postselected wavefunction no longer admits a universal form and depends on the details of initial states, the entanglement entropy exhibits a remarkably universal behavior.
Let the number of sites in subsystem $A_R$ be $l$ with $l\ge L/2$. As shown in the right panel of Fig.~\ref{fig:EE_numeric_main}, the entanglement entropy takes a simple and universal form,
\be
S_{A_R}(l)=N_{\mathbf m}\log 2=(L-l)\log 2,
\ee
which is independent of the specific choice of the measured region $S_P$ in Fig.~\ref{fig:sketch} and remains robust for arbitrary half-filled free-fermion initial states. This observation suggests the existence of a more general organizing principle beyond Theorem~\ref{theorem:2}: even when the post-selected wavefunctions themselves are non-universal for $N_{\mathbf m} < L/2$, they nevertheless encode a universal entanglement structure.

Establishing this universality directly at the wavefunction level is subtle, as the entanglement entropy is a highly nonlinear functional of the state. Developing more powerful analytical tools to address this challenge is therefore left for future work.

\section{Relation between post-measurement probability and entanglement property in the intial state}
\label{Sec:AmpEHRelation}

As shown in Eq.~\eqref{eq:bell_basis_slater}, the probability of obtaining the post-measurement state $|\Psi_{\rm post}\rangle = [\otimes_{i \in A_L}|+\rangle] \otimes [\otimes_{i\in A_R}|-\rangle]$ is given by the product of Slater determinants of the measured region $A_L$ and its complementary $A_R$ as
\begin{equation}
    P = \Big| \Delta_{A_L} \Delta_{A_R} \Big|^2 = \Big| \det [\Phi_0(A_L)] \det [\Phi_0(A_R)] \Big|^2 .
\end{equation}

In this section, we show that this expression can be reformulated directly in terms of the entanglement spectrum of the subsystem $A_L$ in a single layer. This reveals a precise and elegant connection between the probability of obtaining the maximally entangled post-measurement state and the entanglement structure of the initial Gaussian state.

For fermionic Gaussian states, all information about a subsystem $A_L$ is encoded in its single-particle orbital matrix $\Phi_0(A_L)$.  
In particular, the correlation matrix of a subsystem $A_L$ is given by 
\begin{equation}
    C(A_L) = [\Phi_0(A_L)]^\dagger [\Phi_0(A_L)], 
\end{equation}
which is related to the entanglement Hamiltonian through~\cite{Peschel2003}
\begin{equation}\label{eq:def_EH_correlation}
    H_E(A_L) = - \log \rho_{A_L} = - \log \left( [C(A_L)]^{-1} - 1 \right),
\end{equation}
where $\rho_{A_L}$ denotes the reduced density matrix of subsystem $A_L$. 
Using basic determinant identities, we obtain
\begin{equation}
\label{eq:amp_corre}
P = \Big| \det[C(A_L)] \det[C(A_R)] \Big| 
= \prod_i \xi_i^2 = \prod_i \frac{1}{(e^{\varepsilon_i} + 1)^2},
\end{equation}
where ${\xi_i}$ are the eigenvalues of $C(A_L)$ (which coincide with those of $C(A_R)$ since $A_R=[L]\setminus A_L$ and $|A_L|=|A_R|=L/2$), and ${\varepsilon_i}$ are the corresponding entanglement energies.
The index $i=1,\dots,L/2$ runs over the $|A_L|=L/2$ measured sites.
The second equality uses the fact that all eigenvalues are real, and the third equality follows from Eq.~\eqref{eq:def_EH_correlation}.

At half filling, particle-hole symmetry ensures that the correlation-matrix eigenvalues satisfy $\xi_i\in(0,1)$ and appear in pairs $(\xi_i,1-\xi_i)$, corresponding to entanglement levels $(\varepsilon_i,-\varepsilon_i)$. Therefore Eq.~\eqref{eq:amp_corre} simplifies to
\begin{equation}
    P = \prod_i \xi_i (1-\xi_i) = \prod_i \frac{1}{(e^{\varepsilon_i} + 1)(e^{-\varepsilon_i} + 1)}.
\end{equation}
Taking the logarithm, we obtain the compact relation
\begin{equation}\label{Eq:Amp_epslion}
    \log P = - 2 \sum_{i} \log \Big(2 \cosh \frac{\varepsilon_i}{2}\Big),
\end{equation}
which directly connects the half-system entanglement spectrum of the initial state $|\Psi_0\rangle$ to the probability of projecting onto the universally maximally entangled Bell-pair state.

Moreover, each term in Eq.~\eqref{Eq:Amp_epslion} satisfies the bounds
\begin{equation}
    |\epsilon_i| < \log \left(2 \cosh \frac{\varepsilon_i}{2}\right) < |\epsilon_i|+\log(2),
\end{equation}
leading to the rough estimate
\begin{equation}\label{Eq:AmpRoughRelation}
    P \sim \exp \Big( -\sum_i {|\epsilon_i|}\Big),
\end{equation}
which provides a simple and intuitive approximation for the post-measurement probability in terms of the entanglement spectrum in the initial state.

Several explicit examples illustrating the behavior of the probability $P$ are presented in Appendix~\ref{app:amplitude}.

\section{Discussion and Conclusion}
\label{Sec:Conclusion}

In this work, we investigate a universal and maximal form of entanglement swapping induced by Bell measurements on a bilayer system composed of a fermionic Gaussian state and an initially decoupled identical copy, in arbitrary spatial dimensions.
Remarkably, when the measured region comprises exactly half of the total system, a uniform Bell measurement outcome (either $\otimes_i |+\rangle_{i\bar i}$ or $\otimes_i |-\rangle_{i\bar i}$ where ${i,\bar i}$ labels the measured rungs)
drives the post-measurement state to universally factorize into a product of specific local Bell pairs that entangle dual sites across the two layers. As a result, the remaining degrees of freedom exhibit maximal inter-layer entanglement. Importantly, this phenomenon is entirely independent of the microscopic details of the initial state, depending only on its fermionic Gaussian structure.

\medskip

This universality is intrinsic to fermionic systems. As a contrast, in Appendix~\ref{app:spin} we briefly discuss the case of spin systems, where the post-measurement state generally retains explicit dependence on the initial condition prior to the Bell measurements. That is, the universal maximal entanglement originates fundamentally from fermionic antisymmetry, rather than from any fine-tuned property of the initial state.

To further elucidate the fermionic origin of this result, in Appendix~\ref{app:plucker} we present an alternative proof that formulates fermionic Gaussian states as points on the Grassmannian manifold. In this language, the relevant amplitudes are constrained by the classical Pl\"ucker relations, which are well-known consequences of fermionic antisymmetry and Gaussianity. Although more technically involved than the basis-rotation argument presented in the main text, this approach provides a deeper geometric understanding of why Bell measurements of fermionic Gaussian states universally yield a maximally entangled state.

Moreover, since our results apply to the ground states of critical tight-binding models whose low-energy effective theories are described by free fermionic conformal field theories (CFTs), they suggest that existing boundary CFT descriptions of entanglement swapping~\cite{ashida_measurement_bcft_2024} may require substantial modification in the fermionic setting. It would also be highly interesting to explore how these conclusions change in the presence of interactions, where the pre-measurement state is no longer Gaussian. We leave these questions for future work.

\medskip

In the remainder of this section, we discuss two main questions. First, since the probability of obtaining the desired post-measurement state is small, we discuss strategies to overcome this low-probability barrier. Second, we explore how to generalize our Bell-state measurement protocol to more general and broadly applicable settings.

\subsection{Deterministic preparation of postselect state via imaginary time evolution}

For typical initial states such as ground states of local gapped Hamiltonians, the probability of obtaining the desired post-measurement state decays rapidly with system size (see Appendix~\ref{app:amplitude} for details). Several strategies may be employed to enhance this probability. One natural approach is to engineer the entanglement structure of the initial state via unitary driving.
For instance, when the many-body systems illustrated in Fig.~\ref{fig:sketch} are tuned to criticality, there exist universal schemes for reshaping their entanglement structure using various driving protocols, including periodic and quasiperiodic drivings. Early developments along this direction include Refs.~\cite{2018_WenWu,2020Fan_peak,2020PRR_Lapierre,2020_QuasiPeriodic,2020_Han_Wen,2021part1,2022Spart2,2021_general_Lapierre,Fan_2021}. More recently, it has also been proposed that entanglement properties can be effectively tuned through the introduction of a single localized defect~\cite{2025Ryu}.

\medskip

Here, we emphasize a complementary non-unitary approach based on imaginary-time evolution, which has been proposed as a way to overcome the exponential post-selection barrier inherent in post-selection–dependent phenomena~\cite{2025_Yan}.

The general idea is as follows.
Suppose $H$ is a local Hamiltonian acting on a subsystem $A$. We then select a larger region 
$D$ that contains $A$
as a subset.  It can be shown that there exists a unitary operator $U_D$, supported on $D$,
that faithfully reproduces the effect of the imaginary-time evolution $e^{-H\tau}$, up to an error bounded by the correlation between $A$ and the complement of $D$ \cite{2020_Chan}.
That is, this approach works well for the short-range entangled state\footnote{If the initial state has long-range entanglement, such as the ground state of a gapless (critical) Hamiltonian, then the quasi-local simulation of imaginary-time evolution no longer works with exponentially small error, and the construction of a local unitary $U_D$ that mimics imaginary-time evolution breaks down \cite{2020_Chan}.}.

In our setup, for each rung, the Bell state $|+\rangle_{i\bar i}$ is the ground state of 
\be
H_{i\bar i} = 
\Delta (n_i + n_{\bar{i}} - 1)^2
-
t\big( c_i^\dagger c_{\bar{i}} + c_{\bar{i}}^\dagger c_i \big),
\nonumber
\ee
where $n_l = c_l^\dagger c_l$ and $\Delta>t>0$.
Note that if we add a further constrain that the total
number of electrons on this rung is 1, then one can remove the first term in $H_{i\bar i}$. If one wants to 
prepare the Bell state $|-\rangle_{i\bar i}$, one can simply choose $t<0$ in $H_{i\bar i}$.

Next, we choose a buffer region $D$ such that $(i,\bar{i}) \subset D$, 
with the requirement that correlations between the rung $(i,\bar{i})$ and the complement of $D$ are already exponentially suppressed 
in the initial Gaussian state.  

Let
$K_{i\bar i} \equiv e^{-\tau H_{i\bar i}}$
be the imaginary-time evolution that prepares the Bell ground state of $H_{i\bar i}$ 
as $\tau \to \infty$.  
Since $H_{i\bar i}$ is strictly supported on the rung, applying $K_{i\bar i}$ only modifies 
correlations within a light cone controlled by~$\tau$.  
Standard Lieb-Robinson bounds imply that there exists a unitary $U_D$, supported entirely on $D$, 
such that for any operator $O_A$ acting on the rung $A=(i,\bar{i})$,
\begin{align}
K_{i\bar i}^\dagger O_A  K_{i\bar i}
=
U_D^\dagger O_A \, U_D
+ O(\epsilon),
\nonumber
\end{align}
where the approximation error $O(\epsilon)$ is controlled by correlations between $A$ and $C$.
Intuitively, the unitary $U_D$ ``distills'' the Bell state on the rung 
using only degrees of freedom inside $D$, while having negligible effect outside.

In practice, $U_D$ can be constructed explicitly as the time-ordered exponential of a 
quasi-local Hermitian generator,
\begin{equation}
U_D
=
\mathcal{T} \exp\Big[
-i \int_{0}^{\tau} ds\, H^{\mathrm{eff}}_{D}(s)
\Big],
\nonumber
\end{equation}
where $H^{\mathrm{eff}}_{D}(s)$ is obtained by quasi-localizing the Heisenberg-evolved operator 
$H_{i\bar i}(s)$ inside region $D$  (See Ref.\cite{2020_Chan} for details of how to construct 
$H^{\mathrm{eff}}_{D}$ from $H_{i\bar i}$).  
The resulting unitary approximates 
imaginary-time filtering of the Bell subspace with an error that decays exponentially in 
the size of the buffer region.

Applying $U_D$ to the many-body state performs a deterministic and state-dependent step,
$
|\Psi\rangle \mapsto U_D |\Psi\rangle ,
$
which increases the overlap with $|+\rangle_{i\bar i}$ without relying on probabilistic 
post-selection.  
Repeating this for every rung in the set $S_P$ produces, after a finite sequence of 
quasi-local unitaries, a state that is arbitrarily close to the Bell-pair product state
$
\bigotimes_{i\in S_P} |+\rangle_{i\bar i}.
$

In short, replacing post-selection with the above quasi-local unitaries, which is only applied in region $S_P$, produces the same fixed point of the protocol, i.e., the universal Bell-pair product state, but with deterministic success.

\subsection{General projective measurement}

We have shown that maximal entanglement swapping can be achieved by performing projective measurements onto the Bell states in Eq.~\eqref{Bell_+}, namely  $\ket{\pm}_{i\bar i}=(\ket{01}_{i\bar i}\pm\ket{10}_{i\bar i})/\sqrt{2}$.
A natural question then arises: What happens if the projective measurement is not a Bell-state measurement? Do universal features persist in this more general setting?
In this part, we address these questions.

\medskip
For each rung (see Fig.~\ref{fig:sketch}), with the total electron number fixed to unity, we consider a pair of orthonormal states that span the single-occupancy Hilbert space on that rung:
\be
  \label{eq:psi-eps-plus}
|+\rangle^\epsilon_{i\bar i}=
\frac{1}{\sqrt{2}}\Big(\sqrt{1+\epsilon}\,\ket{01}_{i\bar i}
  +\sqrt{1-\epsilon}\,\ket{10}_{i\bar i}\Big),
\ee
and 
\be
  \label{eq:psi-eps-minus}
|-\rangle^\epsilon_{i\bar i}=
\frac{1}{\sqrt{2}}\Big(\sqrt{1-\epsilon}\,\ket{01}_{i\bar i}
  -\sqrt{1+\epsilon}\,\ket{10}_{i\bar i}\Big),
\ee
where $\epsilon \in [-1,1]$. For $\epsilon = 0$, these states reduce to the Bell states employed in the maximal entanglement swapping protocol, while for $\epsilon = \pm 1$ they become the trivial product states $\ket{01}_{i\bar i}$ and $\ket{10}_{i\bar i}$, respectively.

\begin{figure}[t]
    \centering
    \includegraphics[width=0.8\linewidth]{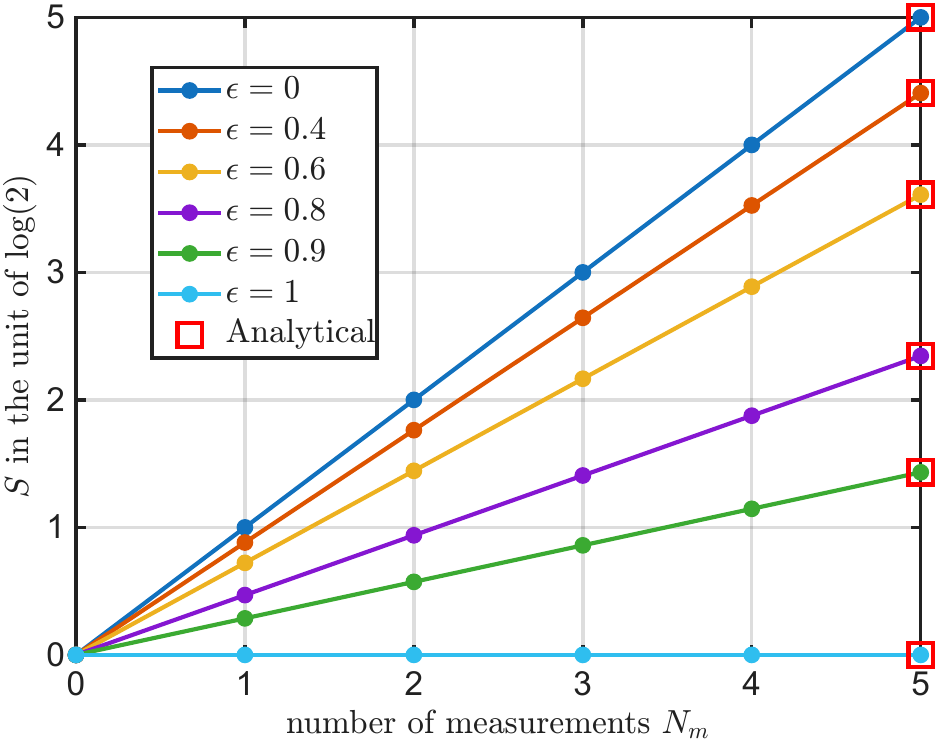}
    \caption{Numerical results for the entanglement entropy of $A_R$ (See Fig.\ref{fig:sketch}) when
    fewer than half of the rungs are measured
    in the imperfect Bell states $|+\rangle_{i\bar i}^\epsilon$. 
    Solid points denote numerical results, and the red squares correspond to analytical results in \eqref{SAR_imperfect} and \eqref{SAR_imperfect2}. The system size is $L=10$.
    }
    \label{fig:ImperfectMeasure}
\end{figure}

We now repeat the measurement protocol described in the previous sections: we measure the region $S_P$ (which is the left half of the total system) shown in Fig.~\ref{fig:sketch} and project each rung onto the state $|+\rangle^{\epsilon}_{i\bar i}$, rather than the ideal Bell state $|+\rangle^{\epsilon=0}_{i\bar i}$. As proven in Sec.~\ref{Sec:Proof}, performing this projective measurement on every rung in $S_P$ yields the post-selected state
\be
\label{Postselect_State_imperfectBell}
    |\Psi_{\text{post}}\rangle = \left[ \otimes_{i \in A_L} |+\rangle^\epsilon_{i\bar i}\, \right] \otimes \left[ \otimes_{i \in A_R} |-\rangle^\epsilon_{i\bar i}\, \right].
\ee
That is, after the post-selection, each rung in the unmeasured region is in the state $|-\rangle^\epsilon_{i\bar i}$ in \eqref{eq:psi-eps-minus}.
This simple analytical structure allows us to compute the entanglement entropy of the subsystem $A_R$ explicitly:
\be
\label{SAR_imperfect}
S_{A_R}(L/2)=\frac{L}{2}\, S^{\,\epsilon}_{\text{rung}},
\ee
where
$L$ is the total number of sites in each layer, and
$S^{\,\epsilon}_{\text{rung}}$ is the entanglement entropy contributed by a single rung 
in the state $|+\rangle_{i\bar i}^\epsilon$,
\be
\label{SAR_imperfect2}
S^{\,\epsilon}_{\text{rung}}=
-\frac{1-\epsilon}{2}\log\Big(\frac{1-\epsilon}{2}\Big)-\frac{1+\epsilon}{2}\log\Big(\frac{1+\epsilon}{2}\Big).
\ee
The analytical results in Eqs.~\eqref{SAR_imperfect} and \eqref{SAR_imperfect2} are shown in Fig.~\ref{fig:ImperfectMeasure}, highlighted by the red squares. In particular, for $\epsilon=0$, we have $S_{A_R}=\frac{L}{2}\log 2$, in agreement with Eq.~\eqref{SA_half_L}.

We further investigate the general case in which the number of projective measurements satisfies $N_{\mathbf m} < L/2$, i.e., fewer than half of the rungs are measured in the basis $\ket{+}^{\epsilon}_{i\bar i}$. Analogous to the Bell-state measurement case (see Fig.~\ref{fig:EE_numeric_main}), our numerical calculations reveal a universal linear growth of the entanglement entropy $S_{A_R}$ as a function of $N_{\mathbf m}$ (as shown in Fig.~\ref{fig:ImperfectMeasure}):
\be
\label{SA_universal_2}
    S_{A_R}(l)=N_{\mathbf m} S^{\,\epsilon}_{\text{rung}} =(L-l)S^{\,\epsilon}_{\text{rung}},
\ee
where $l>L/2$ denotes the number of rungs in the unmeasured region $S_P^c$ shown in Fig.~\ref{fig:sketch}, and $S_{\rm rung}^{\epsilon}$ is given by Eq.~\eqref{SAR_imperfect2}.

We emphasize that for $N_{\mathbf m} < L/2$, the corresponding post-selected state $|\Psi_\text{post}\rangle$ is not universal and depends sensitively on the details of the initial state. This behavior contrasts with the half-system measurement case $N_{\mathbf m} = L/2$, for which 
$|\Psi_\text{post}\rangle$ is universal, as shown in Eq.~\eqref{Postselect_State_imperfectBell}.
Remarkably, despite this lack of universality at the level of the wavefunction, the entanglement entropy 
$S_{A_R}(l)$ in Eq.~\eqref{SA_universal_2} remains universal. 

This suggests the existence of an underlying universal structure encoded in these non-universal post-selected states. We leave a detailed understanding of this intriguing phenomenon for future work.

\begin{acknowledgments}
We thank for interesting discussions with Qihao Cheng, Brian Kennedy, Itamar Kimchi, Zhu-Xi Luo, and Bin Yan. This work is supported by a startup at Georgia Institute of Technology.
\end{acknowledgments}


\bibliography{ref}

\widetext
\appendix


\section{An alternative approach to Theorem 1: Slater determinant and Pl\"ucker relation}
\label{app:plucker}

In this appendix, we provide an alternative proof of Theorem~\ref{corollary:post_measure}, focusing on the half-filling case. This proof employs the well-known Pl\"ucker relation on the Grassmannian $\Gr(N,L)$: the manifold whose points represent free-fermionic states with $N$ occupied orbitals among $L$ available sites. Although this approach is more involved than the basis-rotation method introduced in the main text, it makes the role of antisymmetry explicit and thereby highlights the fermionic nature of the phenomenon.

\subsection{Free fermionic states as points on Grassmannian}

Here, we briefly introduce the concept of the Grassmannian and its connection to free fermionic states for a potentially broader perspective. 

As we have discussed in the main text, a free fermionic state with $L$ orbitals (sites) and $N$ spinless fermions filled can be expanded in the standard computational basis as
\be
|\Psi\rangle = \sum_{A \subset [L], |A| = N} \psi_A \prod_{i \in A} c_{i}^\dagger \vac ,
\ee
where the wavefunction
\be
\psi_A = \Delta_A = \det \Phi([1, \cdots, N], [i_1, \cdots, i_N])
\ee
is known as the Slater determinant for sites $A = [i_1, \cdots, i_N]$ and orbitals $q = [1, \cdots, N]$.

Mathematically, such a state $|\Psi\rangle$ can be viewed as a point on Grassmannnian $\Gr(N, L)$, which is a manifold that parameterizes the set of all $N$-dimensional linear sub-spaces of a $L$-dimensional vector space. From a algebraic point of view, the Grassmannnian $\Gr(N, L)$ is a set of all full-rank $(N \times L)$ matrices with a projection of ${\rm GL}(N)$ that omits all basis transformations of the $N$-dimensional sub-spaces. As we have mentioned, a free fermionic state with $L$ orbitals and $N$ fermions filled (therefore implies particle-number conservation) is fully determined by a $(N \times L)$ matrix $\Phi^{\rm sub}(N, L) = \Phi_{q = 1, \cdots, N}$, with a projection to omits all basis transformations of filled orbitals. This is in line with the definition of a point on $\Gr(N, L)$.

\subsection{Quadratic relation of Pl\"ucker coordinates}

The points on a Grassmannian are identified by the \textit{Pl\"ucker coordinates}, which are homogeneous coordinates~\footnote{Homogenous coordinates, also known as projective coordinates, are coordinates that omits global scale transformations, i.e. $(x_1, \cdots, x_n)$ and $(a \cdot x_1, \cdots, a \cdot x_n)$ represent the same point.} that parameterized by a set of all determinants of the $(N \times N)$ sub-matrices of the $(N \times L)$ matrix (a point on $\Gr(N, L)$). In other words, the coordinates on $\Gr(N, L)$ for identifying a free fermionic state $|\Psi\rangle$ are Slater determinants, as
\be
\left( \Delta_{A_1}, \cdots, \Delta_{A_l} \right) ,
\ee
where $l =\binom{L}{N}$ is the combination number of selecting $N$ from $L$ columns. 

It is easy to see that the Pl\"ucker coordinates are not independent, since the dimension of a Grassmannian is $\dim \Gr(N, L) = N (L - N)$, which is strictly smaller than $\binom{L}{N}$. The constrain on the Pl\"ucker coordinate is given by the following quadratic relation, known as the \textit{Pl\"ucker relation}
\be\label{app_eq:single_column_plucker_relation}
\sum_{l=1}^{N+1} (-1)^l 
\det \left( \Phi^{\text{sub}}_{A \cup \{j_l\}} \right) 
\det \left( \Phi^{\text{sub}}_{B / \{j_l\}} \right) = 0,
\ee
where $\Phi^{\rm sub} \in \Gr(N, L)$, $A = \{i_1, \cdots, i_{N-1}\}$, $B = \{j_1, \cdots, j_{N+1}\}$. This relation can be easily proven by expanding the left determinant along the last column with index $j_l$, and the resummation of the coefficients with the right determinant vanishes (see details in Ref.~\onlinecite{KleimanLaksov}). 
The number of independent such constrains is 
\begin{equation}
\left(\begin{matrix} L \\ N \end{matrix}\right) - 1 - N(L - N) .
\end{equation}
Here the $-1$ term appears because the scaling transformation is quietened: homogeneous coordinates related by a global non-zero scalar factor are equivalent.

As discussed in Ref.~\onlinecite{KleimanLaksov}, there is a more powerful quadratic relation as a generalized form of the Pl\"ucker relation. For a given $\Phi^{\rm sub} \in \Gr(N, L)$ and sets of indices for selecting columns of $\Phi^{\rm sub}$ (occupied sites) $E, F, S \subset [L]$ with size $|E| = k, |F| = N - k - 1, |S| = N + 1$, the following relation holds
\begin{equation}\label{app_eq:general_plucker}
\sum_{ \substack{ C \subset S, |C| = N - k \\ D = S / C } } \sgn(C \cup D) \Delta_{E \cup C} \Delta_{D \cup F} = 0 .
\end{equation}
It is obvious that, by setting
\be
E = A, \ F = \emptyset, \ S = B ,
\ee
the general quadratic relation (later referred to as the generalized Pl\"ucker relation) in Eq.~\eqref{app_eq:general_plucker} reduces to the one shown in Eq.~\eqref{app_eq:single_column_plucker_relation}. Similar to the relation in Eq.~\eqref{app_eq:single_column_plucker_relation}, here the generalized quadratic relation can be derived by using Laplace expansion of the determinants, but along multiple columns. This relation holds for any $(N \times L)$ full-rank matrix.

\subsection{The post-measurement wavefunction in real-space representation}

We are interested in the wavefunction of basis that contain the component of 
\begin{equation}
\otimes_{i \in A_L} \frac{1}{\sqrt{2}} \left( |01\rangle_{i \overline{i}} + |10\rangle_{i \overline{i}} \right) ,
\end{equation}
where $A_L \subset [L] = \{1, \dots, L\}$ and $|A_L| = L/2$. This is equivalent to the post-measurement wavefunction after performing $\frac{L}{2} = N$ Bell measurements to the set of pairs $(i, \overline{i}), i \in A_L$. 

Without losing generality, we let $A_L = [L/2] = \{ 1, \dots, L/2 \}$. After performing $\frac{L}{2} = N$ Bell measurements on the left half of the bilayer system (note that the sites can be arranged in arbitrary order, so it is not restricted to the physical position of ``left half''), the state of the total system becomes
\begin{equation}
	\begin{aligned}
		| \Psi_{\rm{post}} \rangle = \left[ \otimes_{i=1}^{N} \frac{1}{\sqrt{2}} \left( |01\rangle_{i \overline{i}} + |10\rangle_{i \overline{i}} \right) \right] \otimes | \Psi_{\rm{remain}} \rangle 
		= \prod_{i=1}^{N} \frac{1}{\sqrt{2}} (c_i^\dagger + c_{\overline{i}}^\dagger) | {\rm{vac}} \rangle_{\rm left}
		\otimes | \Psi_{\rm{remain}} \rangle , 
	\end{aligned}
\end{equation}
where we have assumed that $| \Psi_{\rm{remain}} \rangle$ is normalized, i.e. there is a non-zero probability of projecting the left half to a product of Bell states. 

\subsubsection{Notations for defining the fermionic ordering}

Before expanding the wavefunction, let us define the site ordering by classifying the occupied sites (in the computational basis) into the following sets
\begin{equation}
	\begin{aligned}
		A_L & = \{ i_1, \cdots , i_a \} , \ A_R = \{ i_{a+1}, \cdots , i_N \}, \ i_a \leq N
		\\ 
		\overline{B}_L & = \{ \overline{j}_1, \cdots , \overline{j}_b \} , \ \overline{B}_R = \{ \overline{j}_{b+1}, \cdots , \overline{j}_N \}, \ \overline{j}_b \leq N
	\end{aligned}
\end{equation}
with assuming ascending order of the site indices
\begin{equation}
	i_1 < \cdots < i_m , \quad \overline{j}_1 < \cdots < \overline{j}_n .
\end{equation}
For convenience, we introduce the following notations
\begin{equation}
	A = A_L \cup A_R, \quad \overline{B} = \overline{B}_L \cup \overline{B}_R
\end{equation}
to represent the sets of occupied sites in the upper and lower layers, respectively. We will also use the notation $B_L, B_R, B$ to represent the occupied sites in the lower layer, defined as $\overline{B}_L = \{ \overline{j}, j \in B_L \}$, $\overline{B}_R = \{ \overline{j}, j \in B_R \}$ and $B = B_L \cup B_R$, for the purpose of comparing the site indices between $A$ and $\overline{B}$. Unless otherwise specified, the union operation $A \cup B$ is defined as joining the two set $A$ and $B$ with ordering.

It is important to notice that the size of the above introduced four sets, i.e. the occupation in the post-measurement state, is restricted to a specific pattern as
\be\label{app_eq:particle_number_set_constrain}
|A_L| = |B_R| = a, \quad |A_R| = |B_L| = b = N - a, 
\ee
This is nothing but the consequence of the half-filling setting, for which the Bell measurements does not change the total particle number. 
Moreover, since the left half of the bilayer has been projected to the product of Bell pairs, we have 
\begin{equation}\label{app_eq:left_set_constrain}
	A_L \cup B_L = { \textstyle \left[ N \right] }, A_L \cap B_L = \emptyset 
\end{equation}
as another constraint of allowed computational basis. Here $\left[ N \right]$ represents the list of $[1, \cdots, N]$. The basis, along with any occupations other than the constraints in the Eqs.~\eqref{app_eq:particle_number_set_constrain}~\&~\eqref{app_eq:left_set_constrain}, has a vanishing wavefunction in the post-measurement state. This property will be useful in the proof of Theorem~\ref{corollary:post_measure} at half-filling.

\subsubsection{Explicit form of the post-measurement wavefunction}

Under the above-introduced notations, we expand the post-measurement state as
\begin{equation}
		| \Psi_{\rm{post}} \rangle  = 
		\sum_{A_L, B_L} \psi^{{ {\rm{Bell}}, +}}_{A_L, B_L} \prod_{\alpha=1}^{a} c_{i_\alpha}^\dagger \prod_{\beta =1}^{N - a} c_{\overline{j}_\beta}^\dagger | {\rm{vac}} \rangle_{A_L \cup \overline{B}_L} 
		\otimes \sum_{A_R, B_R} \psi_{A_R, B_R}^{\rm remain} \prod_{\zeta=1}^{N-a} c_{i_\zeta}^\dagger \prod_{\eta=1}^{a} c_{\overline{j}_\eta}^\dagger  | {\rm vac} \rangle_{A_R \cup \overline{B}_R} 
\end{equation}
where we have changed the site ordering from rung-by-rung to layer-by-layer. This will induce a sign in the wavefunction. In particular, under the rearranged site ordering, the wavefunction of the measured left-half part is
\begin{equation}
	\psi_{A_L, \overline{B}_L}^{{\rm Bell}, +} = \begin{cases}
		0 & A_L \cap B_L \neq \emptyset 
		\\ 
		\sgn \left( \sigma (A_L \cup B_L) \right)
		\left( \frac{1}{\sqrt{2}} \right)^{N} & A_L \cap B_L = \emptyset .
	\end{cases}
\end{equation}
Here $\sigma$ represents the permutation of the joint set $A_L \cup B_L$ to ascending order, and $\sgn(\sigma)$ is the sign change induced by the permutation of fermionic operators. 

The main subject of this work is the wavefunction of the remaining unmeasured part in the post-measurement state. Solving $\psi^{\rm remain}$ is equivalent to finding the wavefunction in the pre-measurement state $|\Psi\rangle_{\rm double} = |\Psi_0\rangle \otimes | \overline{\Psi}_0 \rangle$ for those basis with occupations that satisfy the constrains in Eqs.~\eqref{app_eq:particle_number_set_constrain}~\&~\eqref{app_eq:left_set_constrain}. To achieve this, we rearrange the site ordering in the wavefunction representations from Eq.~\eqref{eq:rep_double_psi_1} to a new basis that separates the left and right halves of the bilayer. Under this transformation, the representation of a computational basis changes as
	\be
	\begin{aligned}
		|r\rangle & = \prod_{i \in A} c_i^\dagger | {\rm vac} \rangle_{A_L \cup A_R}  \otimes \prod_{j \in B} c_{\overline{j}}^\dagger | {\rm vac} \rangle_{\overline{B}_L \cup \overline{B}_R} 
		= (-1)^{|A_R| \cdot |B_L|} \prod_{i_L \in A_L} c_{i_L}^\dagger \prod_{j_L \in B_L} c_{\overline{j}_L}^\dagger 
		\prod_{i_R \in A_R} c_{i_R}^\dagger \prod_{j_R \in B_R} c_{\overline{j}_R}^\dagger \vac
	\end{aligned}
	\ee
	which leads to the following expression of the wavefunction of the remaining unmeasured part
	\be\label{eq:psi_remain_reexpress}
	\begin{aligned}
		\psi^{\rm remain}_{A_R, B_R} 
		& = \sum_{ {A_L \subset [N], |B_L| = [N] - |A_L|} }  (-1)^{|A_R| \cdot |B_L|} \sgn\left( \sigma(A_L \cup B_L) \right) \Delta_{A_L \cup A_R} \Delta_{B_L \cup B_R} 
	\end{aligned}
	\ee
upto a global scale factor. This expression is helpful for converting the quadratic product of determinants to the wavefunction, which will be used for the proof of Theorem~\ref{corollary:post_measure} via the P\"ucker relation. 

\subsection{The approach to Theorem 1 via P\"ucker relation}

Here, to prove the statement of Theorem~\ref{corollary:post_measure} at half-filling, we adopt the generalized Pl\"ucker relation that introduced in Eq.~\eqref{app_eq:general_plucker}. 

Now let us recall the statement of Theorem~\ref{corollary:post_measure}, which states that the remaining unmeasured part of the post-measurement state has the following form
\begin{equation}
		| \Psi \rangle_{\rm remain} = \prod_{i=N}^{2N} \left[ \frac{1}{\sqrt{2}} (c_i^\dagger - c_{\overline{i}}^\dagger)  \right] \vac_{A_R \cup \overline{B}_R} 
		= \sum_{ \substack{A_R \subset [N, 2N] \\ B_R = [N, 2N] / A_R} } 
		\psi^{\rm remain, p}_{ A_R, B_R } \prod_{i \in A_R} c_i^\dagger
		\prod_{j \in [N, 2N]} c_{\overline{j}}^\dagger  \vac_{A_R \cup \overline{B}_R} 
\end{equation}
where $[N, 2N]$ represents the list of $[N, N + 1, \cdots, 2N]$. Here the wavefunction is 
\begin{equation}\label{eq:propose_remain_psi}
	\begin{aligned}
		\psi^{\rm remain, p} = \left( \frac{1}{\sqrt{2}} \right)^{N} (-1)^{|B_R|} \sgn \left( \sigma( A_R \cup \overline{B}_R ) \right) ,
	\end{aligned}
\end{equation}
where the permutation $\sigma$ sends the joint set $A_R \cup \overline{B}_R$ to a set with ascending order. If there is a pair of $i$ and $\overline{i}$, $\sigma$ send them to the order of $\{ i, \overline{i} \}$.

We prove Theorem~\ref{corollary:post_measure} by separating it to two Lemmas, which are easily to be shown by using the generalized quadratic relation of determinants in Eq.~\eqref{app_eq:general_plucker}. 

\begin{applemma}\label{app_lemma:1}
	The wavefunctions of the remaining unmeasured part in Eq.~\eqref{eq:psi_remain_reexpress} vanish, i.e. $\psi^{\rm remain}_{A_R, B_R} = 0$, when $A_R \cap B_R \neq \emptyset$. 
\end{applemma}
\noindent \textit{Proof. }
Assume there is at least one overlapping element $b_0 = (B_R)_\beta \in A_R \cap B_R$, then by setting $E = A_R, F = B_R / \{ b_0 \}, S = A_L \cup B_L \cup \{ b_0 \}$. Under this setting, the generalized quadratic relation of Pl\"ucker relation in Eq.~\eqref{app_eq:general_plucker} becomes 
\begin{equation}\label{app_eq:lemma_1}
	\begin{aligned}
		0 & = \sum_{ \substack{C \subset S, |C| = N - |E| \\ D = S / C} } \sgn \left( \sigma( C \cup D ) \right) \Delta_{E \cup C} \Delta_{D \cup F} 
		\\ & 
		= \sum_{ \substack{C = A_L \cup \{ b_0 \}  \\ D = S / C = B_L } } \sgn \left( \sigma(A_L \cup \{b_0\} \cup B_L) \right) \Delta_{A_R \cup A_L \cup \{b_0\}} \Delta_{B_L \cup B_R / \{b_0\}}
		\\ & \quad 
		+ \sum_{\substack{ C = A_L \subset S \\ D = S / C = B_L \cup \{b_0\} }} \sgn \left( \sigma( A_L \cup B_L \cup \{ b_0 \} ) \right) \Delta_{A_R \cup A_L} \Delta_{B_L \cup \{b_0\} \cup B_R / \{b_0\}}
		\\ & 
		= \sum_{A_L \in [N], |A_L| = N - |A_R|} (-1)^{|B_L|} \sgn\left( \sigma(A_L \cup B_L) \right) (-1)^{|A_R| \cdot |A_L|} \Delta_{A_L \cup A_R \cup \{b_0\}} \Delta_{B_L \cup B_R / \{b_0\}} 
		\\ & \quad 
		+ \sum_{A_L \in [N], |A_L| = N - |A_R|} \sgn\left( \sigma(A_L \cup B_L) \right) (-1)^{|A_R| \cdot |A_L|} \Delta_{A_L \cup A_R} (-1)^{\beta-1} \Delta_{B_L \cup B_R}
	\end{aligned}
\end{equation}
where in the last line we have used the fact that
\be
\sgn\left( \sigma(A_L \cup \{b_0\} \cup B_L \right) = (-1)^{|B_L|} \sgn\left( \sigma(A_L \cup B_L) \right) , 
\quad
\sgn\left( \sigma(A_L \cup B_L \cup \{b_0\}) \right) = \sgn\left( \sigma(A_L \cup B_L) \right)
\ee
and
\be
\sigma(A_L \cup B_L ) = [N], \quad |A_L| = |B_R| = N - |A_R| = N - |B_L| .
\ee
By comparing with the representation of remaining part wavefunction $\psi^{\rm remain}$ in Eq.~\eqref{eq:psi_remain_reexpress}, we have
\begin{equation}
	\begin{aligned}
		\sum_{A_L \in [N], |A_L| = N - |A_R|} \sgn\left( \sigma(A_L \cup B_L) \right) \Delta_{A_L \cup A_R \cup \{b_0\}} \Delta_{B_L \cup B_R / \{b_0\}} = & (-1)^{(|A_R|+1)^2} \psi^{\rm remain}_{A_R \cup \{b_0\}, B_R / \{ b_0 \}}  
		\\ 
		\sum_{A_L \in [N], |A_L| = N - |A_R|} \sgn\left( \sigma(A_L \cup B_L) \right) \Delta_{A_L \cup A_R} (-1)^{\beta-1} \Delta_{B_L \cup B_R} = & (-1)^{|A_R|^2} \psi^{\rm remain}_{A_R, B_R} ,
	\end{aligned}
\end{equation}
which interprets the quadratic product of determinants to the wavefunction. 
Combing with the relation between $|A_L|, |A_R|, |B_L|, |B_R|$, the above quadratic relation then becomes
\begin{equation}
	0 = (-1)^{|A_R| \cdot |A_L|} \Big[ \psi^{\rm remain}_{A_R \cup \{b_0\}, B_R / \{ b_0 \}}  + (-1)^{\beta-1} \psi^{\rm remain}_{A_R, B_R}   \Big] 
\end{equation}
which leads to 
\begin{equation}\label{eq:lemma_1}
	\psi^{\rm remain}_{A_R \cup \{b_0\}, B_R / \{ b_0 \}}  + (-1)^{\beta-1} \psi^{\rm remain}_{A_R, B_R} = 0
\end{equation}

It is obvious that the first term in the last line of Eq.~\eqref{eq:lemma_1} vanishes, since $A_R$ and $\{b_0\}$ have overlapping index $b_0$. Then we have 
\begin{equation}
	\psi^{\rm remain}_{A_R, B_R} = 0
\end{equation}
when $A_R \cap B_R \neq \emptyset$. $\qed$

The physical meaning of the statement of Lemma~\ref{app_lemma:1} is that, the remaining unmeasured part of the post-measurement state can be written in terms of product of wavefunctions of local pairs of upper-layer site $i$ and its dual $\overline{i}$ in the lower layer, i.e. there is no coupling between those local pairs $\{ i, \overline{i} \}$. 

\begin{applemma}\label{app_lemma:2}
	The wavefunctions of the remaining unmeasured part in Eq.~\eqref{eq:psi_remain_reexpress} have an identical amplitude and an appreciated sign factor as the proposed form in Eq.~\eqref{eq:propose_remain_psi}, when $A_R \cap B_R = \emptyset$. In particular, they have the following expression
	\begin{equation}
		\psi^{\rm remain}_{A_R, B_R} = (-1)^{|B_R|} \sgn \left( \sigma(A_R \cup \overline{B}_R) \right) \Delta_{[N]} \Delta_{[N, 2N]}
	\end{equation}
	upto a global phase factor.
\end{applemma}
\noindent \textit{Proof. } By setting $E = A_R, F = B_R / \{ b_0 \}, S = A_L \cup B_L \cup \{ b_0 \}$, $b_0 = (B_R)_\beta \in B_R$, similar to the derivation of Lemma~\ref{app_lemma:1}, the generalized quadratic relation of Pl\"ucker coordinates in Eq.~\eqref{app_eq:general_plucker} becomes 
\begin{equation}\label{app_eq:lemma_2}
	\begin{aligned}
		0 & = \sum_{ \substack{A_L \in [N], |A_L| = N - |A_R| \\ B_L = [N] / A_L} } \sgn \left( \sigma(A_L \cup B_L \cup \{b_0\}) \right) 
		\Delta_{A_R \cup A_L} \Delta_{B_L \cup {b_0} \cup B_R / \{b_0\}}
		\\ & \quad 
		+ \sum_{ \substack{A_L \in [N], |A_L| = N - |A_R| \\ B_L = [N] / A_L} } \sgn \left( \sigma(A_L \cup \{b_0\} \cup B_L) \right) \Delta_{A_R \cup A_L \cup \{b_0\}} \Delta_{B_L \cup B_R / \{b_0\}} 
		\\ & 
		= \sum_{ \substack{A_L \in [N], |A_L| = N - |A_R| \\ B_L = [N] / A_L} } \sgn \left( \sigma(A_L \cup B_L) \right) (-1)^{|A_R| \cdot |A_L|} (-1)^{\beta-1} \Delta_{A_R \cup A_L} \Delta_{B_L \cup B_R} 
		\\ & \quad 
		+ \sum_{ \substack{A_L \in [N], |A_L| = N - |A_R| \\ B_L = [N] / A_L} } \sgn \left( \sigma(A_L \cup B_L) \right) (-1)^{|B_L|} (-1)^{|A_R| \cdot |A_L|} \Delta_{A_R \cup A_L} \Delta_{A_R \cup A_L \cup \{b_0\}} \Delta_{B_L \cup B_R / \{b_0\}} 
		\\ & 
		= (-1)^{|A_R| \cdot |A_L|} (-1)^{|A_R|^2} (-1)^{\beta - 1} \psi^{\rm remain}_{A_R, B_R} 
		+ (-1)^{|A_R| \cdot |A_L|} (-1)^{|B_L|} (-1)^{(|A_R|+1)^2} \psi^{\rm remain}_{A_R \cup \{b_0\}, B_R / \{b_0\}} ,
	\end{aligned}
\end{equation}
which gives 
\be\label{eq:proof_lemma_2}
(-1)^{\beta - 1} \psi^{\rm remain}_{A_R, B_R} - \psi^{\rm remain}_{A_R \cup \{b_0\}, B_R / \{b_0\}} = 0
\ee
where we have used the representation of the remaining part wavefunction in Eq.~\eqref{eq:psi_remain_reexpress} for interpreting the quadratic product of determinants as the wavefunction. 

This means that the amplitude of the wavefunction $\psi^{\rm remain}$ is identical for given sets of occupation (given computational basis for the remaining part of the post-measurement state) with removing an occupied site from upper to lower layer, vice versa. Starting with the case of $A_R = [N, 2N], B_R = \emptyset$, i.e. the basis where all fermions occupied on the upper layer, and repeating the operation of removing an occupied site from $A_R$ (upper layer) to $B_R$ (lower layer), one can construct all possible combinations of $A_R \in [N, 2N], B_R = [N, 2N] / A_R$. Eq.~\eqref{eq:proof_lemma_2} then leads to the conclusion that all $\psi^{\rm remain}$ have the same amplitude when $A_R \cup B_R = [N, 2N], A_R \cap B_R = 0$. The sign change induced by the removing operation is obviously $(-1)^{|B_R|} \sgn \left( \sigma(A_R \cup B_R) \right)$. To see this, one can consider that the removing operation is always applied for the largest site index in $A_R$, such that $b_0 = (B_R)_{\beta=1}$ is always the smallest index in $B_R$. In this case, for each removing operation, there is a single sign change and eventually leads to $(-1)^{|B_R|}$. Applying a permutation to the standard ordering in the computational basis then gives another sign factor of $\sigma(A_R \cup B_R)$. $\qed$

\vspace{2ex}
\noindent \textbf{Proof of the statement of Theorem~\ref{corollary:post_measure} at half-filling.} The statement of Theorem~\ref{corollary:post_measure} at half-filling is an immediate consequence of Lemmas~\ref{app_lemma:1} \&~\ref{app_lemma:2}.  $\qed$ 

Note that the above two Lemmas are proven under the condition that the probability of projecting the touched half part to the product of Bell states $\otimes_{i=1}^{N} |+\rangle_{i \overline{i}}$ is non-zero, which was introduced when assuming the form of the post-measurement state in Eq.~\eqref{eq:post_state}. As we have stated in the main text, this condition is satisfied by choosing the initial state $|\Psi_0\rangle$ to be non-separable for any bipartition in real space. 

\subsection{An alternative proof to Proposition~\ref{proposition}}

In the main text, we prove Proposition~\ref{proposition} directly from the Slater determinants of $|\Psi_0\rangle$ that gives post-measurement state. Here, we present an alternative proof from another angle. In particular, we consider an equivalent statement of Proposition~\ref{proposition} by converting it to the Bell measurement problem as stated in the following:
\begin{appprop}
	The post-measurement state is non-normalizable, i.e. the probability of projecting to the product of Bell states is zero, if the state $|\Psi_0\rangle$ can be factorized to a simple product of two parts of the total system, i.e. the state can be written as
	\begin{equation}
		| \Psi_0 \rangle = | \Psi_G \rangle \otimes | \Psi_{G^c} \rangle, \quad G \cap G^c = \emptyset, \ G \cup G^c = [L] . 
	\end{equation}
	Unless both $|G|$ and $|G^c|$ are even, and the number of Bell measurements performed in $|G|$ and $|G^c|$ are $\frac{|G|}{2}$ and $\frac{|G^c|}{2}$, respectively. 
\end{appprop}
\noindent \textit{Proof. } Suppose $|G| = \frac{L}{2} + k, |G^c| = \frac{L}{2} - k, k \in [0, \frac{L}{2}]$. For the site set $G^c$, at most ${\rm round}(\frac{L}{4} - \frac{k}{2})$ Bell measurements are allowed to host a non-zero probability. This follows from the fact that the maximal entanglement supported by a pure state in a Hilbert space of dimension $2^{|G^c|}$ is $\frac{|G^c|}{2} \log 2$. Then the remaining ${\rm ceil}(\frac{L}{4} + \frac{k}{2})$ measurements must be applied on $G$. If $|G|$ is odd, then ${\rm ceil}(\frac{L}{4} + \frac{k}{2}) > \frac{|G|}{2}$, which yields a zero probability. If $|G|$ is even, then by Theorem~\ref{corollary:post_measure}, the only case that can have a non-zero probability is when the number of Bell measurements performed in $|G|$ and $|G^c|$ are $\frac{|G|}{2}$ and $\frac{|G^c|}{2}$, respectively. $\qed$

\section{Vanishing post-measurement wavefunction from non-separable initial states $|\Psi_0\rangle$}
\label{app:non_separable}

In the main text, we have shown that spatially separable states $|\Psi_0\rangle$ are generally trivial, in the sense that the wavefunction vanishes after projecting a half of the doubled state $|\Psi_{\rm double}\rangle = |\Psi_0\rangle \otimes |\overline{\Psi}_0\rangle$, where $|\overline{\Psi}_0\rangle = |\Psi_0\rangle$ is an identical copy, to a product of $|+\rangle_{i} \equiv |+\rangle_{i\overline{i}}$. 
However, the converse does not hold, i.e. separable states are only one class of trivial states, and not all trivial states are separable. To see this, note that for separable states the orbital matrix can be permuted into block-diagonal form, with all off-block entries equal to zero. By contrast, a vanishing Slater determinant can still have non-zero entries throughout: it merely requires that the matrix of selected orbitals is not full rank.

Here, to make the statement more intuitive, we present a minimal example of a non-separable $|\Psi_0\rangle$ that yields a vanishing post-measurement wavefunction. The construction is simple. Recall that the post-measurement wavefunction is non-vanishing if $|\Psi_0\rangle$ has non-zero amplitudes for both the computational basis $\prod_{i \in A_L} c_i^\dagger | {\rm vac} \rangle$ and $\prod_{i \in A_R} c_i^\dagger | {\rm vac} \rangle$, where $A_L$ is the region of upper-layer sites on which Bell measurements applied and $A_R = [L] / A_L$ is its complementary. Thus, to construct an example of non-separable but trivial $|\Psi_0\rangle$, we just need to let its orbital matrix be rank-deficient for the set $A_L$. 

Let us first consider that the initial state $|\Psi_0\rangle$ lives only on two sites $i=1,2$. In this case, at half-filling the general form of the state is
\begin{equation}
|\Psi_0\rangle = \alpha | 10 \rangle + \beta | 01 \rangle, \quad |\alpha|^2 + |\beta|^2 = 1
\end{equation}
and the orbital matrix is
\begin{equation}
\Phi_0 = \left(\begin{matrix}
\alpha & \beta 
\end{matrix}\right) , \quad \text{with} \quad \Phi_0 \left(\begin{matrix} c_1^\dagger \\ c_2^\dagger  \end{matrix}\right) = \alpha c_1^\dagger + \beta c_2^\dagger = b_1^\dagger . 
\end{equation} 
In this case, the post-measurement state has vanishing wavefunction, if and only if at least one of the two parameters $\alpha$ and $\beta$ equals to zero, i.e. the state $|\Psi_0\rangle$ is a product state in real space. 

Therefore, the minimal example of a non-separable but trivial state at least contains 2 fermions and lives on a 4-site chain. 
Consider for example the following orbital matrix
\begin{equation}
\Phi_0 = \left(\begin{matrix}
\frac{1}{2}  & \frac{1}{2} & \frac{1}{2} & \frac{1}{2} \\ 
\frac{\sqrt{6}}{6} & \frac{\sqrt{6}}{6} & 0 & -\frac{\sqrt{6}}{3} 
\end{matrix}\right) 
\end{equation}
which gives
\begin{equation}
|\Psi_0\rangle = \frac{1}{2} (c_1^\dagger + c_2^\dagger + c_3^\dagger + c_4^\dagger) \frac{\sqrt{6}}{6} (c_1^\dagger + c_2^\dagger - 2 c_4^\dagger) | {\rm vac} \rangle_{1234} = - \frac{\sqrt{6}}{12} ( c_1^\dagger c_3^\dagger + 3 c_1^\dagger c_4^\dagger + c_2^\dagger c_3^\dagger + 3 c_2^\dagger c_4^\dagger + 2 c_3^\dagger c_4^\dagger ) | {\rm vac} \rangle_{1234}.
\end{equation}
This is a well-defined free fermionic state with conserved particle number. It is non-separable since the orbital matrix cannot be written in terms of a direct sum of orbital matrices for any real-space bipartition. 
However, by choosing $A_L = \{1,2\}$ and $A_R = \{1,2,3,4\} / A_L = \{3,4\}$, the post-measurement state vanishes since 
\begin{equation}
\Delta_{A_L} = \det [\Phi_{0}(A_L)] = \det\left(\begin{matrix} \frac{1}{2} & \frac{1}{2} \\ \frac{\sqrt{6}}{6} & \frac{\sqrt{6}}{6} \end{matrix}\right) = 0 . 
\end{equation}

\section{The initial-state dependence for spin models}
\label{app:spin}

Let us first consider the case that the initial state lives on a 2-site chain, which has the following general form
\begin{equation}
|\Psi_0\rangle = \alpha | 10 \rangle_{12} + \beta | 01 \rangle_{12}, \quad |\alpha|^2 + |\beta|^2 = 1.
\end{equation}
Here we only consider the basis with total $S = 0$, as an analog to the particle number conservation of fermions. Note that the basis states are bosonic instead of fermionic, and we have used $0$ and $1$ to represent spin-down and spin-up, respectively. The corresponding doubled state is 
\begin{equation}
\begin{aligned}
|\Psi_{\rm double}\rangle & = |\Psi_0\rangle \otimes |\overline{\Psi}_0\rangle = (\alpha | 10 \rangle_{12} + \beta | 01 \rangle_{12}) \otimes (\alpha | 10 \rangle_{\overline{1} \overline{2}} + \beta | 01 \rangle_{\overline{1} \overline{2}}) \\ 
& = \alpha^2 |1010\rangle_{12 \overline{1} \overline{2}} + \beta^2 |0101\rangle_{12 \overline{1} \overline{2}} + \alpha \beta ( |1001\rangle_{12 \overline{1} \overline{2}} + |0110\rangle_{12 \overline{1} \overline{2}} ).
\end{aligned}
\end{equation}
After performing Bell measurement $P_+ = |+\rangle \langle+|$ on the pair of sites $\{1, \overline{1}\}$, we leave only
\begin{equation}
|\Psi_{\rm post}\rangle = |+\rangle_{1, \overline{1}} \otimes |+\rangle_{2, \overline{2}} ,
\end{equation}
i.e. the post-measurement state is a product of two Bell pairs, similar to the fermionic case but slightly different on the measurement outcome ($|+\rangle \otimes |-\rangle \to |+\rangle \otimes |+\rangle$).

The minimal example for an initial-state dependent post-measurement state is given by a 4-site chain, which has the following general form
\begin{equation}
|\Psi_0\rangle = (\alpha_1 \sigma_1^+ + \alpha_2 \sigma_2^+ + \alpha_3 \sigma_3^+ + \alpha_4 \sigma_4^+)
(\beta_1 \sigma_1^+ + \beta_2 \sigma_2^+ + \beta_3 \sigma_3^+ + \beta_4 \sigma_4^+) |0000\rangle
\end{equation}
i.e.
\begin{equation}
|\Psi_0\rangle = c_1 |1100\rangle + c_2 |1010\rangle + c_3 |1001\rangle + c_4 |0110\rangle + c_5 |0101\rangle + c_6 |0011\rangle
\end{equation}
with
\begin{equation}
\begin{aligned}
	c_1 = \alpha_1\beta_2+\alpha_2\beta_1, \quad c_2 = \alpha_1\beta_3+\alpha_3\beta_1, \quad c_3 = \alpha_1\beta_4+\alpha_4\beta_1, 
	\\ 
	\quad c_4 = \alpha_2\beta_3+\alpha_3\beta_2, \quad c_5 = \alpha_2\beta_4+\alpha_4\beta_2, \quad c_6 = \alpha_3\beta_4+\alpha_4\beta_3,
\end{aligned}
\end{equation}
as the constrain that the state is free. Here all the positive sign in the coefficient will become negative when considering the state is fermionic, e.g. $c_1 =  \alpha_1\beta_2+\alpha_2\beta_1 \to  \alpha_1\beta_2-\alpha_2\beta_1$. 

Now, let us perform Bell measurements $P_+$ on pairs of sites $\{1,\overline{1}\}$ and $\{2,\overline{2}\}$, the doubled state $|\Psi_{\rm double}\rangle = |\Psi_0\rangle \otimes |\overline{\Psi}_0\rangle$. To clear see the difference between fermionic and bosonic states, below we will explicit leave the sign change due to exchanging (anti-)symmetric particles. The post-measurement state has the following expression
\begin{equation}
|\Psi_{\rm post}\rangle = \prod_{i=1}^{2} \otimes |+\rangle_{i, \overline{i}} \otimes ( {}_{1, \overline{1}} \langle+| \otimes {}_{2, \overline{2}} \langle+| \otimes \mathbf{1}_{34\overline{3}\overline{4}} ) |\Psi_0\rangle_{1234} \otimes |\overline{\Psi}_0\rangle_{\overline{1234}} 
= \prod_{i=1}^{2} \otimes |+\rangle_{i, \overline{i}} \otimes |\Psi_{\rm remain} \rangle
\end{equation}
where
\begin{equation}
\begin{aligned}
|\Psi_{\rm remain}\rangle
&= \frac{c_1 c_6}{2}( |1100\rangle_{34\overline{34}} + |0011\rangle_{34\overline{34}} )
+ \frac{c_3 c_5 \pm c_3 c_5}{2} |0101\rangle_{34\overline{34}} 
+ \frac{c_2 c_4 \pm c_2 c_4}{2} |1010\rangle_{34\overline{34}} 
\\ & \quad 
+ \frac{(c_2 c_5 \pm c_3 c_4)}{2} ( |0110\rangle_{34\overline{34}} \pm |1001\rangle_{34\overline{34}} )
\end{aligned}
\end{equation}
with the sign $\pm$ denotes the statistics: $+$ for bosoinic and $-$ for fermionic. In case the state is fermionic, it becomes
\begin{equation}
\begin{aligned}
|\Psi_{\rm remain}\rangle & =  |1100\rangle_{34\overline{34}} + |0011\rangle_{34\overline{34}} +  |0110\rangle_{34\overline{34}} - |1001\rangle_{34\overline{34}} 
\\ & = (|110\rangle_{34\overline{3}} + |011\rangle_{34\overline{3}}) \otimes |0\rangle_{\overline{4}} 
+ ( |001\rangle_{34\overline{3}} - |100\rangle_{34\overline{3}} ) \otimes |1\rangle_{\overline{4}} 
\\ & = (|10\rangle_{3\overline{3}} - |01\rangle_{3\overline{3}}) \otimes |10\rangle_{4\overline{4}} 
+ (|01\rangle_{3\overline{3}} - |10\rangle_{3\overline{3}}) \otimes |01\rangle_{4\overline{4}} 
= |-\rangle_{3\overline{3}} \otimes |-\rangle_{4\overline{4}}
\end{aligned}
\end{equation}
upto a global normalization factor (here we have used $c_1c_6 = c_2c_5 - c_3c_4 \neq 0$). However, in the bosonic case, the post-measurement state is clearly initial-state dependent. The difference is only a consequence of the symmetric and anti-symmetric statistics of the bosons and fermions.

\section{Projection to other Bell states}
\label{appendix:OtherBell}

Here we present a minimal example with $L=2$ to illustrate what happens when projecting onto Bell states other than $|\pm\rangle$. Consider the doubled initial state
\be
|\Psi\rangle_{\rm double} = | \Psi_0 \rangle \otimes | \overline{\Psi}_0 \rangle, \qquad 
| \Psi_0 \rangle = (\alpha c_1^\dagger + \beta c_2^\dagger) \vac_{12}, \quad 
| \overline{\Psi}_0 \rangle = (\alpha c_{\overline{1}}^\dagger + \beta c_{\overline{2}}^\dagger) \vac_{\overline{1} \overline{2}} ,
\ee
and the Bell measurement is given by the following projective operator
\be
P^{\uparrow}_{1 \overline{1}} = | \uparrow \rangle_{1 \overline{1}} \ {}_{1 \overline{1}}\langle \uparrow |, 
\qquad 
| \uparrow \rangle_{1 \overline{1}} = \frac{1}{\sqrt{2}} (c_1^\dagger c_{\overline{1}}^\dagger + {\mathbf{1}}_{1 \overline{1}}) \vac_{1 \overline{1}} = \frac{1}{\sqrt{2}} (|00\rangle + |11\rangle)_{1 \overline{1}},
\ee
where $\mathbf{1}_{1\overline{1}}$ denotes the identity operator.

After applying this projective measurement, the post-selected state becomes
\be
| \Psi_{\rm post} \rangle = \frac{ P^{\uparrow}_{1 \overline{1}} | \Psi \rangle_{\rm double} }{\norm{ P^{\uparrow}_{1 \overline{1}} | \Psi \rangle_{\rm double} }} = \frac{1}{\sqrt{2}} (|00\rangle + |11\rangle)_{1 \overline{1}} \otimes \frac{1}{\sqrt{|\alpha^2|^2 + |\beta^2|^2}} (\alpha^2 |00\rangle + \beta^2|11\rangle)_{2 \overline{2}} .
\ee
It is clear that the post-measurement state now depends sensitively on the coefficients of the initial state $|\Psi_0\rangle$. In particular, the resulting two-layer state no longer exhibits the universal, maximally entangled Bell-pair structure that arises when projecting onto $|\pm\rangle$.

\section{Imperfect copy of the initial state $|\Psi_0\rangle$}
\label{app:imperfect}

In the main text, we focused on the case in which the bilayer system consists of an initial state $|\Psi_0\rangle$ and its identical copy $|\overline{\Psi}_0\rangle = |\Psi_0\rangle$. In this section, we instead consider the situation where the two layers differ slightly, i.e., $|\overline{\Psi}_0\rangle$ is a perturbed version of $|\Psi_0\rangle$. Specifically, we take $|\Psi_0\rangle$ to be the ground state of a fermionic chain with mass parameter $m_0$,
\begin{equation}
	H_0 = - \frac{1}{2} \sum_{i} c_i^\dagger c_{i+1} + h.c.+m_0\sum_i (-1)^i c_i^\dagger c_{i} ,
\end{equation}
and set $|\overline{\Psi}_0\rangle = |\Psi_{\rm perturb}\rangle$ to be the ground state of $H_0$ in the presence of an additional perturbation,
\begin{equation}
	H_{\rm perturb} = \delta m\sum_i (-1)^i  c_i^\dagger c_{i} ,
\end{equation}
where $\delta m$ denotes a small shift of the mass parameter.

We perform a direct numerical calculation to study the resulting post-measurement state. Throughout, we fix $m_0 = 0.3$, perform the half-system Bell measurement, and analyze the post-selected state on the unmeasured region $S_P^c = A_R \cup B_R$. To characterize this state, we compute two quantities as functions of the system size $L$ and the perturbation strength $\delta m$:
(i) the entanglement entropy between $A_R$ and $B_R$, and
(ii) the overlap fidelity
\[
F=\bigl|\langle \Phi_{\rm Ideal}|\Psi_{\rm post}\rangle\bigr|^2,
\]
where $|\Phi_{\rm Ideal}\rangle= \otimes_{i \in A_R} |-\rangle_{i}$ is the ideal maximally entangled Bell-pair product state that is predicted in our main theory and $|\Psi_{\rm post}\rangle$ is the perturbed post-measurement state in $S_P^c$.

\begin{figure}[htbp]
    \centering
    \subfigure{
        \includegraphics[width=0.39\linewidth]{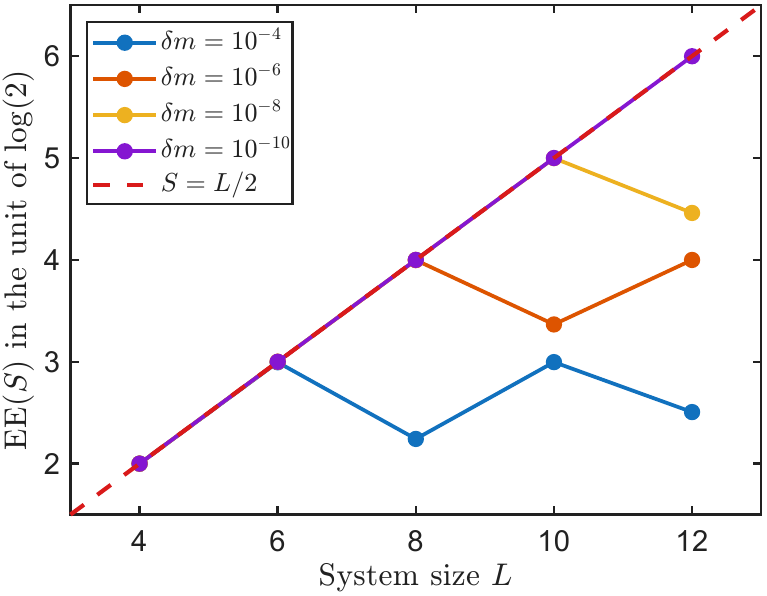}
    }
    \hspace{-10pt} 
    \subfigure{
        \includegraphics[width=0.4\linewidth]{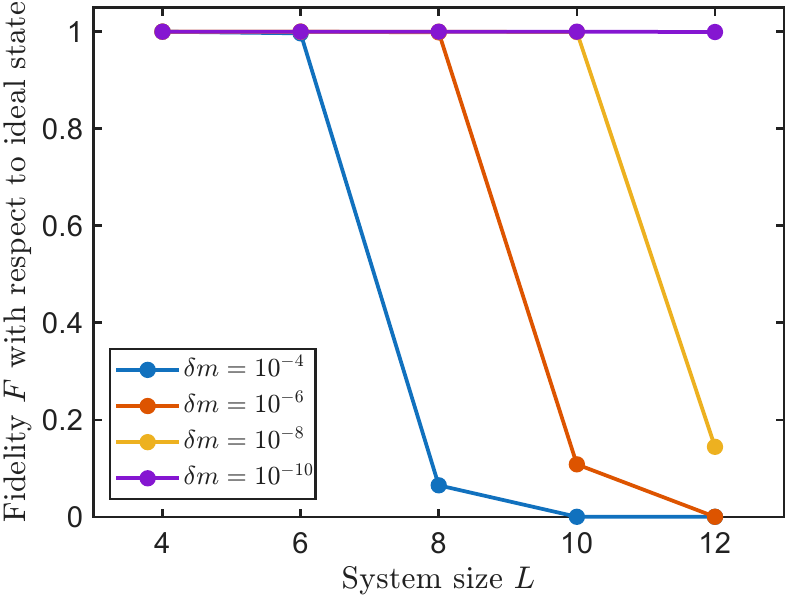}
    }
    \caption{ Numerical results for half-system Bell measurements in a bilayer with an imperfect copy.  Left: EE between $A_R$ and $B_R$ (in units of $\log 2$) as a function of the single-layer length $L$. Right: overlap fidelity $F=|\langle \Phi_{\rm Ideal}|\Psi_{\rm post}\rangle|^2$ as a function of $L$.}
    \label{fig:PerturbationPlot}
\end{figure}

The results (see Fig.~\ref{fig:PerturbationPlot}) demonstrate that, for sufficiently small $\delta m$, the protocol continues to prepare a nearly maximally entangled state on the unmeasured degrees of freedom. However, as the system size increases, deviations from the ideal Bell-pair structure become progressively more pronounced. In other words, achieving (nearly) maximally entangled Bell states in larger systems requires correspondingly smaller perturbations in the copied layer.

\section{Dependence of the probability of post-measurement state on the system size $L$ and the initial mass $m$}
\label{app:amplitude}

In this appendix, we focus on a class of typical initial states given by the ground states of local free-fermion Hamiltonians, and analyze how the probability of obtaining the post-measurement state depends on the system size $L$ and the initial mass parameter $m$.
\begin{figure}\centering
	\includegraphics[width=0.9\columnwidth]{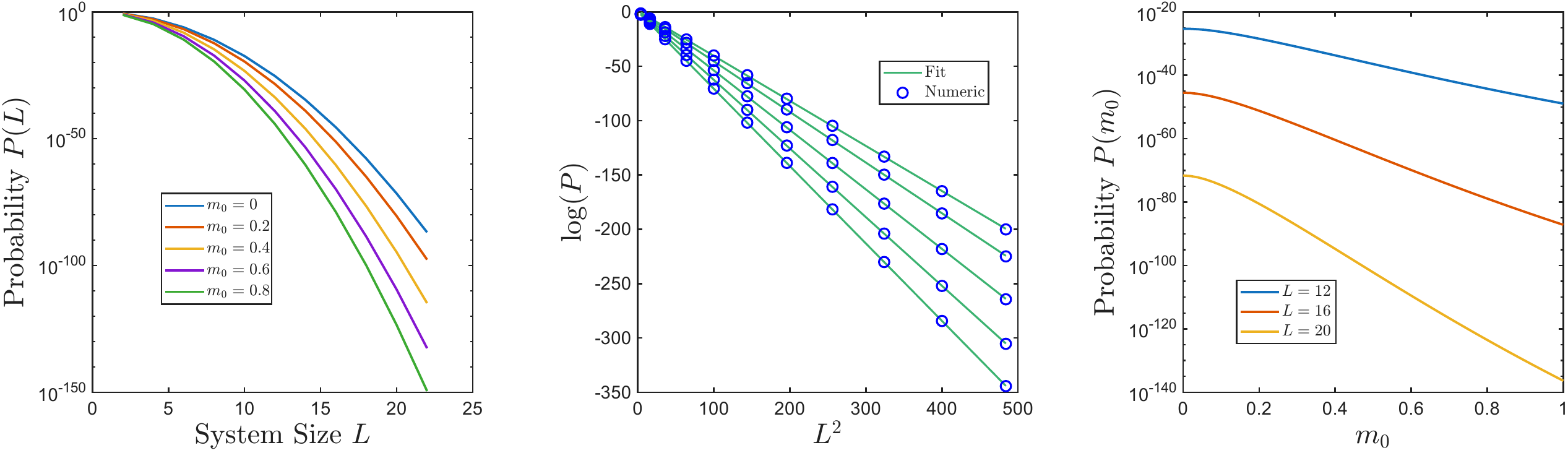}
	\caption{
		\label{fig:AmpChange}
        Numerical results of projection probability $P = |\Delta_{A_L} \Delta_{A_R}|^2$, where the initial states are chosen as the ground states of $H_0$ in Eq.\eqref{H0_appendixF}.
Left: $P$ plotted as a function of the single-chain system size $L$ for various values of the initial mass $m_0$.
Middle: Replot of the left panel with the horizontal axis $L^2$ and the vertical axis $\log P$. Blue circles denote the numerical data, while solid green lines show linear fits of the form $\log P \simeq \alpha + \beta L^2$.
Right: $P$ plotted as a function of the initial mass $m_0$ for several values of the system size $L$.
	}
\end{figure}

Here we take the initial state $|\Psi_0\rangle$ to be the half-filled ground state of a free Dirac fermion lattice model with mass parameter $m$ and open boundary conditions,
\begin{equation}
\label{H0_appendixF}
H_0 = - \frac{1}{2} \Big( \sum_{i=1}^{L-1} c_i^\dagger c_{i+1} + h.c. \Big) + m_0 \sum_{i} (-1)^i \, c_{i}^\dagger c_{i},
\end{equation}
where the last term corresponds to a staggered potential that generates the Dirac mass.

As we have shown in the main text, the probability of obtaining the postselection state is given by $P=|\Delta_{A_L}\Delta_{A_R}|^2$, where $\Delta_S$ denotes the Slater determinant for the set of occupied sites $S$, $A_L$ is the set of measured sites in the upper layer, and $A_R = [L] / A_L$ is the complementary of $A_L$ in the upper layer.

In Fig.~\ref{fig:AmpChange}, we numerically evaluate the projection probability $P$ for various values of the system size $L$ and the initial mass parameter $m_0$ in Eq.~\eqref{H0_appendixF}. The results show that $P$ decays exponentially with $L^2$.
This scaling behavior can be understood as follows. In Sec.~\ref{Sec:AmpEHRelation}, we established a connection between the post-measurement probability $P$ and the entanglement spectrum ${\varepsilon_i}$,
\be
\label{P_appendix}
P\sim \exp\Big(-\sum_i |\epsilon_i| \Big), 
\ee
which is expected to hold when the entanglement energies $|\varepsilon_i|$ are sufficiently large. According to Eisler’s work~\cite{eisler2025bisognano}, the entanglement spectrum of the present model exhibits an equally spaced structure,
\begin{equation}\label{Eq:StateEntgSpec}
        \epsilon_i=(2i-\frac{1}{2})\epsilon, \quad \epsilon=2\pi\frac{K(\kappa')}{K(\kappa)}, \quad i\in \mathbb Z,
\end{equation}
where $K$ is the complete elliptic integral of the first kind, and $\kappa,\kappa'$ is related to the mass by
\begin{equation}
    \kappa = \frac{1}{\sqrt{1+m_0^2}}~,~\kappa'=\frac{m_0}{\sqrt{1+m_0^2}}.
\end{equation}
Substituting Eq.~\eqref{Eq:StateEntgSpec} into Eq.~\eqref{P_appendix} and retaining only the $L$-dependent contribution, we obtain the estimate
\begin{equation}
    P\sim \exp(-(L^2+C)\epsilon),
\end{equation}
where $C$ is an $L$-independent constant, and $\epsilon$ depends solely on the mass parameter $m_0$ and may therefore be treated as a constant for fixed $m_0$.
Consistent with this analysis, a linear fit of $\log P$ versus $L^2$, shown in the middle panel of Fig.~\ref{fig:AmpChange}, exhibits excellent agreement with the numerical results.
Figure~\ref{fig:AmpChange} (right panel) also shows that the projection probability $P$ decays exponentially with the mass parameter $m_0$. This behavior is expected, as increasing the mass suppresses correlations and reduces the entanglement present in the initial state.

\end{document}